\documentclass[%
 aip,
 jcp,
 amsmath,amssymb,
 reprint,
]{revtex4-2}

\usepackage{graphicx}
\usepackage{dcolumn}
\usepackage{bm}

\usepackage[utf8]{inputenc}
\usepackage[T1]{fontenc}
\usepackage{mathptmx}
\usepackage{etoolbox}
\usepackage{xcolor}
\usepackage{booktabs}
\usepackage{multirow}
\usepackage{url}
\usepackage{hyperref}
\hypersetup{
colorlinks=true,
linkcolor=blue,
citecolor=blue,
filecolor=green,
urlcolor=blue,
}

\makeatletter
\def\@email#1#2{%
 \endgroup
 \patchcmd{\titleblock@produce}
  {\frontmatter@RRAPformat}
  {\frontmatter@RRAPformat{\produce@RRAP{*#1\href{mailto:#2}{#2}}}\frontmatter@RRAPformat}
  {}{}
}%
\makeatother
\begin{document}


\title[]{Triplet adiabatic potential energy surfaces for the O$(^3\mathrm{P})$ + C$_2(\mathrm{X}\,^1\Sigma_\mathrm{g}^+, \mathrm{a}\,^3\Pi_\mathrm{u})$	  reaction and quasiclassical trajectory dynamics}
\author{Anzhela Veselinova--Marinova}
\altaffiliation[Present address: ]{Univ. Rennes, CNRS, IPR (Institut de Physique de Rennes), UMR 6251, Rennes F-35000, France}
\affiliation{Departamento de Química Física, Facultad de Ciencias Químicas, Universidad de Salamanca, 37008 Salamanca, Spain}
\author{Susana G\'omez--Carrasco}
\affiliation{Departamento de Química Física, Facultad de Ciencias Químicas, Universidad de Salamanca, 37008 Salamanca, Spain}
\author{Lola Gonz\'alez--S\'anchez}
\affiliation{Departamento de Química Física, Facultad de Ciencias Químicas, Universidad de Salamanca, 37008 Salamanca, Spain}
\author{Alexandre Zanchet}
\affiliation{Instituto de Física Fundamental , CSIC , Serrano 123 , Madrid, 28006, Spain.}
\author{Pablo G. Jambrina}
\affiliation{Departamento de Química Física, Facultad de Ciencias Químicas, Universidad de Salamanca, 37008 Salamanca, Spain}
\author{Cristina Sanz--Sanz}
\affiliation{Departamento de Química Física Aplicada (Unidad Asociada de I+D+i al CSIC), Universidad Autonoma de Madrid, Modulo 14 , Madrid 28049, Spain}

\date{\today}

\begin{abstract}
We present three global adiabatic potential energy surfaces (PESs) for the O$(^3\mathrm{P})$ + C$_2(\mathrm{X}\,^1\Sigma_\mathrm{g}^+, \mathrm{a}\,^3\Pi_\mathrm{u}) \to$ CO$(\mathrm{X}\,^1\Sigma^+)$ + C$(^3\mathrm{P})$ reaction, computed at the icMRCI+Q/aug-cc-pVTZ level. This reaction is relevant to combustion chemistry, atmospheric processes, and astrochemistry. Since the energy difference between the $^1\Sigma_\mathrm{g}^+$ and $^3\Pi_\mathrm{u}$ states of C$_2$ is smaller than the zero-point energy, both states mix upon interaction with the approaching O atom. Consequently, the PESs exhibit a double-well topography in the reactant valley. Beyond this feature, the three PESs display markedly different topographies. The $1\,^3A''$ PES features a deep minimum 7.16 eV below the O + C$_2$ entrance channel. In contrast, the $2\,^3A''$ and $1\,^3A'$ surfaces possess shallower collinear wells ($\approx$ 5.7 eV) and strong angular anisotropy arising from avoided crossings with higher-lying triplet states. The analytical fits reproduce the \textit{ab initio} data with root-mean-square (RMS) deviations of 0.022--0.028 eV over an energy interval exceeding 13 eV. Quasiclassical trajectory (QCT) calculations reveal higher reactivity on the barrierless $1\,^3A''$ surface compared to the $2\,^3A''$ PES, where an angle-dependent entrance barrier reduces the cross sections. Furthermore, cross sections for reactions initiated from C$_2(\mathrm{X}\,^1\Sigma_\mathrm{g}^+)$ and C$_2(\mathrm{a}\,^3\Pi_\mathrm{u})$ are remarkably similar for the three PESs, regardless of the initial rovibrational state.
\end{abstract}

\maketitle

\section{Introduction}\label{sec:introduction}

Atomic oxygen is a dominant species in low Earth orbit (LEO) owing to the low dissociation energy of O$_2$, making O + carbon interactions relevant to spacecraft materials\cite{troya_intrevphyschem_2004}. Beyond this context, the C$_2$ radical and its reactions with oxygen play important roles in several high-energy environments, including combustion chemistry\cite{paramo_jphyschema_2008,gu_jphyschema_2006,pasternack_chemphys_1981,barovavski_jchemphys_1977,reisler_chemphys_1980}, atmospheric processes\cite{paramo_jphyschema_2006,guo_aipconfproc_2006,combi_astrophysj_1997,federman_astrophysj_1989,federman_astrohpysj_1988,green_annyrevphyschem_1981}, and the shock layers generated during hypersonic re-entry into planetary atmospheres\cite{kruse_sympintcomust_1998}. In fuel-rich flames, C$_2$ acts as a key intermediate influencing soot formation and PAH growth\cite{barovavski_jchemphys_1977,reisler_jchemphys_1980,mangir_jchemphys_1980}, while in re-entry conditions, collisions with atomic oxygen affect the chemical evolution of carbon-containing atmospheres\cite{troya_intrevphyschem_2004,kruse_sympintcomust_1998}. C$_2$ is also a widely observed molecule in astrophysical environments, including cold interstellar and circumstellar regions, comets, and stellar atmospheres, as well as in other environments such as plasmas and hydrocarbon flames\cite{paramo_jphyschema_2006,kazmierczak_mnras_2010,welty_mnras_2013,cox_aa_2014,Loison:NNRAS114}, further underscoring the importance of characterizing its reactivity with atomic oxygen. In this context, the O + C$_2$ $\to$ CO + C reaction is also included in astrochemical databases\cite{smith_mnras_2004}  as a possible neutral-neutral process, although its behaviour at low temperatures remains unknown due to the absence of measurements in that regime.

A distinctive feature of the C$_2$ molecule is the very small energy splitting between its two lowest electronic states, $\mathrm{X}\,^1\Sigma_\mathrm{g}^+$ and $\mathrm{a}\,^3\Pi_\mathrm{u}$, which are separated by only 718~cm$^{-1}$ (0.089~eV)\cite{kirby_jchemphys_1979,veset_canjphys_1975}. Furthermore, the equilibrium bond length of the C$_2(^3\Pi_\mathrm{u})$ state is slightly larger than that of the C$_2(^1\Sigma_\mathrm{g}^+)$ state. Combined with the small energy gap, this implies that the $^3\Pi_\mathrm{u}$ state becomes the electronic ground state at slightly stretched C--C distances. Notably, this crossing occurs at a bond length shorter than the outer classical turning point of the C$_2(^1\Sigma_\mathrm{g}^+, v=0, j=0)$ state.

Although the energy gap between the C$_2(^1\Sigma_\mathrm{g}^+)$ and C$_2(^3\Pi_\mathrm{u})$ states ensures that the singlet state dominates in the low energy limit, most laboratory production techniques generate C$_2$ in both states simultaneously and do not allow sufficient electronic relaxation to reach thermal equilibrium. As a result, experimental samples invariably contain mixed singlet/triplet populations, complicating the interpretation of kinetic data for reactions involving C$_2$. This complexity is reflected in the limited measurements available for the O + C$_2$ system, where existing experiments probe only room-temperature or high-temperature conditions. Becker \textit{et al.}~\cite{becker_zphyschem_2000} measured a rate coefficient of $(9.8 \pm 1.0) \times 10^{-11}$~cm$^3$~s$^{-1}$ for the O$(^3\mathrm{P})$ + C$_2(^3\Pi_\mathrm{u})$ reaction at room temperature. Kruse and Roth~\cite{kruse_sympintcomust_1998} investigated the O$(^3\mathrm{P})$ + C$_2(^1\Sigma_\mathrm{g}^+)$ channel in shock-tube experiments between 2750 and 3950~K, obtaining $k = 1.26 \times 10^{-9} \exp(-6250/T)$~cm$^3$~s$^{-1}$. Additionally, Federman and Huntress~\cite{federman_astrophysj_1989} reported a value of $4 \times 10^{-11}$~cm$^3$~s$^{-1}$ at room temperature. Consequently, the temperature dependence of this reaction outside of these specific regimes remains uncharacterized.

\begin{figure}[ht!]
  \centering
  \includegraphics[width=1.0\linewidth]{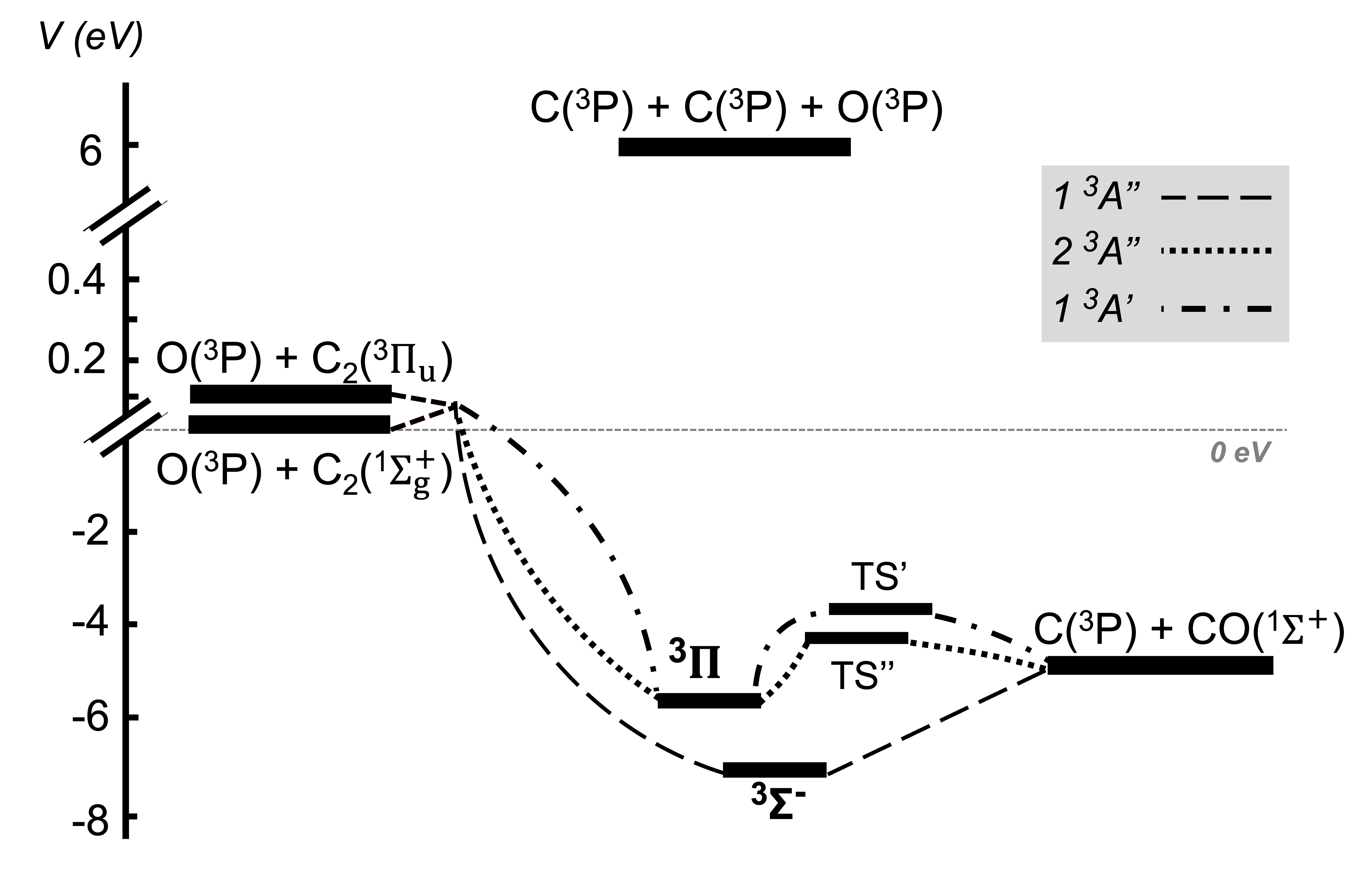}
\caption{
Correlation diagram for the three lowest triplet PESs, labeled in C$_s$ symmetry ($1\,^3A''$, $2\,^3A''$ and $1\,^3A'$), illustrating their adiabatic correlation with the O$(^3\mathrm{P})$ + C$_2(\mathrm{X}\,^1\Sigma_\mathrm{g}^+, \mathrm{a}\,^3\Pi_\mathrm{u})$ reactant channels and the CO$(\mathrm{X}\,^1\Sigma^+)$ + C$(^3\mathrm{P})$ product asymptote. The stationary points characterized in this work are shown, where each PES displays a global minimum at linear geometry.}
  \label{fig:diagram}
\end{figure}

From a theoretical perspective, to the best of our knowledge three studies have addressed the O$(^3\mathrm{P})$ + C$_2$ reaction to date. Ray, Saha, and Schatz~\cite{schatz_jphyschemc_2012} performed on-the-fly nonadiabatic quasiclassical trajectory (QCT) calculations using the trajectory surface-hopping formalism. Their electronic structure calculations were carried out using the B3LYP functional and the 6-31G(d,p) basis set. Stationary points for both the O$(^3\mathrm{P})$ + C$_2(^1\Sigma_\mathrm{g}^+)$ and O$(^3\mathrm{P})$ + C$_2(^3\Pi_\mathrm{u})$ channels were optimized at the DFT level and compared with CCSD(T)/cc-pVTZ single-point energies. Their QCT calculations, performed at hyperthermal collision energies between 1 and 5~eV, predicted that the CO + C cross section for the O$(^3\mathrm{P})$ + C$_2(^1\Sigma_\mathrm{g}^+)$ reaction is roughly three times larger than that for O$(^3\mathrm{P})$ + C$_2(^3\Pi_\mathrm{u})$. However, the reliance on single-reference DFT and on-the-fly energies limits the accuracy and level of detail that can be achieved for the complex electronic structure of this system. A subsequent study by Wang \textit{et al.}~\cite{wang_chemphys_2019} reported a detailed characterization of the singlet manifold for the O$(^3\mathrm{P})$ + C$_2(^3\Pi_\mathrm{u})$ $\to$ CO$(\mathrm{X}\,^1\Sigma^+)$ + C$(^1\mathrm{D})$ reaction. Using MRCI calculations and more than 50,000 \textit{ab initio} points, they constructed adiabatic PESs for the three lowest singlet states, fitted via three-dimensional B-spline interpolation. That study analyzed one- and two-dimensional cuts of the surfaces, identifying linear and bent intermediates as well as avoided crossings between the lowest states and a higher-lying singlet state. However, the PESs were not provided in an analytical form, and the study was strictly limited to the singlet states correlating with the C$(^1\mathrm{D})$ product asymptote. As depicted in Fig.~\ref{fig:diagram}, this is not the ground electronic state of the product channel; the ground-state CO$(\mathrm{X}\,^1\Sigma^+)$ + C$(^3\mathrm{P})$ asymptote is exclusively accessible via the triplet manifold.

Furthermore, Huang \textit{et al.}~\cite{zang_jchemphys_2015} developed a full-dimensional PES for the $1\,^3A''$ state of the CCO system, which correlates with the O$(^3\mathrm{P})$ + C$_2(^1\Sigma_\mathrm{g}^+)$ $\to$ CO$(\mathrm{X}\,^1\Sigma^+)$ + C$(^3\mathrm{P})$ reaction. Their surface was fitted with a permutation-invariant polynomial neural network (PIP-NN) using 9293 MRCI+Q/aug-cc-pVTZ \textit{ab initio} energies. The resulting PES reproduces the linear minimum of the state---located 7.10 eV below the O + C$_2(^1\Sigma_\mathrm{g}^+)$ entrance channel at $r_{\mathrm{CC}}=2.627\,a_0$ and $r_{\mathrm{CO}}=2.186\,a_0$---and yields a transition state whose geometry and energy closely follow those reported by Ray \textit{et al.}~\cite{schatz_jphyschemc_2012}. The authors also performed QCT calculations for C$_2(^1\Sigma_\mathrm{g}^+, v=0\text{--}10, j=0\text{--}100)$ over collision energies of 0.1--6~eV. They reported state-resolved cross sections that reproduced the DFT direct-dynamics results of Ray \textit{et al.}~\cite{schatz_jphyschemc_2012} and thermal rate coefficients in good agreement with the experimental data of Kruse and Roth~\cite{kruse_sympintcomust_1998}. Nevertheless, this computational study presents three significant limitations: (i) it completely neglects the C$_2(^3\Pi_\mathrm{u})$ state, which is nearly degenerate with the ground C$_2(^1\Sigma_\mathrm{g}^+)$ state; (ii) the cross sections calculated as a function of the collision energy ($E_{\mathrm{col}}$) exhibit the characteristic behavior of an activated reaction---namely, the presence of an energy threshold and an increasing cross section at low $E_{\mathrm{col}}$---which is unphysical given the true barrierless nature of the entrance channel; and (iii) the study only includes one of the three triplet PESs that correlate with the relevant reactant and product asymptotes.

In this paper  we carried out a computational study of the O$(^3\mathrm{P})$ + C$_2$ $\to$ CO$(^1\Sigma^+)$ + C$(^3\mathrm{P})$ reaction. To this end, we characterize the three global triplet PESs (one $1\,^3A'$ and two $^3A''$ states) that correlate adiabatically with the relevant atom--diatom asymptotes: O$(^3\mathrm{P})$ + C$_2(^1\Sigma_\mathrm{g}^+)$, O$(^3\mathrm{P})$ + C$_2(^3\Pi_\mathrm{u})$, and CO$(\mathrm{X}\,^1\Sigma^+)$ + C$(^3\mathrm{P})$. As shown in the correlation diagram of Fig.~\ref{fig:diagram}, both reactant channels correlate with the same three triplet PESs that lead to the ground product channel. It may seem counterintuitive that the same adiabatic PES correlates with two different electronic states in the reactant asymptote, but this is a direct consequence of the change in energetic ordering of the states as a function of the internuclear distance $r_{\mathrm{CC}}$. At shorter $r_{\mathrm{CC}}$ values, the three PESs correlate with the C$_2(^1\Sigma_\mathrm{g}^+)$ state, whereas at larger $r_{\mathrm{CC}}$ values---but still below the outer turning point of the $v=0, j=0$ level---they correlate with the C$_2(^3\Pi_\mathrm{u})$ state. Consequently, in the asymptotic O+C$_2$ region, these PESs exhibit a double-well structure along the $r_{\mathrm{CC}}$ coordinate that resembles   the potential for the umbrella motion in NH$_3$. This behavior further reflects the complex electronic structure of C$_2$, for which double-well potentials have also been reported for the $2 ^1\Pi_u$ and $2 ^1\Sigma^+_u$ excited states of the isolated diatom\cite{Kyle:JCP22}.

The fact that both the C$_2(^1\Sigma_\mathrm{g}^+)$ and C$_2(^3\Pi_\mathrm{u})$ states correlate with the $1\,^3A''$, $2\,^3A''$, and $1\,^3A'$ surfaces makes dynamical studies more complex. Ideally, one would separate the contributions of the two initial states by constructing a fully diabatic representation of the relevant PESs. However, due to the high density of electronic states in this system, such a procedure is currently computationally unaffordable. As an alternative, Rocha \textit{et al.}~\cite{OctavioSC2} restricted their calculations to the equilibrium internuclear distance of C$_2(^1\Sigma_\mathrm{g}^+)$. This geometrical constraint allowed them to treat the C$_2(^1\Sigma_\mathrm{g}^+)$ and C$_2(^3\Pi_\mathrm{u})$ states separately and subsequently perform kinetic calculations using the adiabatic statistical (AS) method, which does not require explicitly considering the vibrational motion of C$_2$. However, fixing the internuclear distance is not an option when performing full-dimensional reactive scattering calculations.

This complication was overlooked in the full-dimensional calculations performed for the O + C$_2(^1\Sigma_\mathrm{g}^+)$ reaction in Ref.~\citenum{zang_jchemphys_2015}. An inspection of that PES suggests that the authors  computed the \textit{ab initio} points in the interaction region and forced the surface to match the C$_2(^1\Sigma_\mathrm{g}^+)$ asymptote. Interestingly, this near-degeneracy issue does not arise for the analogous reaction producing excited carbon, O$(^3\mathrm{P})$ + C$_2(^3\Pi_\mathrm{u})$ $\to$ CO$(\mathrm{X}\,^1\Sigma^+)$ + C$(^1\mathrm{D})$, because there are no PESs that correlate adiabatically with that specific product asymptote and the C$_2(^1\Sigma_\mathrm{g}^+)$ entrance channel.

The paper is organized as follows. Section \ref{sec:methods} details the computational methodology, including the \textit{ab initio} electronic structure calculations, the analytical fitting of the PESs, and the setup for the quasiclassical trajectory (QCT) dynamical calculations. The results and discussion are presented in Section \ref{sec:results}, which is divided into two subsections: the topological analysis of the  fitted PESs and the outcomes of the QCT calculations (namely, cumulative reaction probabilities and cross sections). Finally, the main conclusions of this work are summarized in Section \ref{sec:conclusions}.

\section{Computational methodology}\label{sec:methods}
\subsection{\textit{Ab initio} electronic calculations}

Multiple electronic states are required to properly describe the O$(^3\mathrm{P})$ + C$_2 \to$ CO$(^1\Sigma^+)$ + C$(^3\mathrm{P})$ reaction. Table~\ref{tabcorrdiag} lists the lowest electronic states correlating with the reactant and product asymptotes. These states are classified according to the C$_{\infty v}$ and C$_{s}$ point groups. The interaction between O$(^3\mathrm{P})$ and C$_2(\mathrm{X}\,^1\Sigma_\mathrm{g}^+)$ generates two triplet states that correlate with the $^{3}\Sigma^{-}$ and $^{3}\Pi$ terms in C$_{\infty v}$ symmetry. In bent geometries (C$_{s}$ symmetry), the $^{3}\Sigma^{-}$ term becomes a $^3A''$ state, while the $^{3}\Pi$ term splits into one $^3A'$ and one $^3A''$ state. The system complexity increases significantly, however, for the interaction of O$(^3\mathrm{P})$ with C$_2(^3\Pi_\mathrm{u})$. In C$_s$ symmetry, these fragments yield six singlet, six triplet, and six quintet states (three $A'$ and three $A''$ for each multiplicity). In C$_{\infty v}$ symmetry, these states correlate with the $^{1,3,5}\Sigma^{+}$, $^{1,3,5}\Sigma^{-}$, $^{1,3,5}\Pi$, and $^{1,3,5}\Delta$ terms. The $\Sigma^{+}$ and $\Sigma^{-}$ states yield one $A'$ and one $A''$ state, respectively, whereas the $\Pi$ and $\Delta$ states each split into one $A'$ and one $A''$ state (see Table~\ref{tabcorrdiag}).

The triplet manifold arising from the O$(^3\mathrm{P})$ + C$_2(^1\Sigma_\mathrm{g}^+)$ entrance channel also defines the C$(^3\mathrm{P})$ + CO$(^1\Sigma^+)$ product asymptote, which adiabatically correlates to the same $^3\Sigma^-$ and $^3\Pi$ electronic states. Thus, this manifold fully characterizes the ground reactant and product channels. Conversely, only a subset of the manifold originating from the O$(^3\mathrm{P})$ + C$_2(^3\Pi_{\mathrm{u}})$ channel correlates adiabatically with these three triplet states. The remaining triplet states correlate with the CO$(\mathrm{a}\,^3\Pi)$ + C$(^3\mathrm{P})$ asymptote, which lies more than 1.0 eV above the entrance channel. The singlet and quintet states from the O$(^3\mathrm{P})$ + C$_2(^3\Pi_{\mathrm{u}})$ channel correlate instead with product asymptotes involving CO$(^1\Sigma^+)$ and electronically excited C atoms, as well as the C$(^3\mathrm{P})$ + CO$(^3\Pi)$ asymptote.

\begin{table*}[ht!]
\centering
\begingroup
\setlength{\tabcolsep}{18pt} 
\renewcommand{\arraystretch}{1.5} 
\begin{tabular}{l|c|c|l}
\hline
\multicolumn{1}{c|}{\textbf{Asymptote}}
& \textbf{Multiplicity}
& \textbf{C$_{\infty v}$}
& \multicolumn{1}{c}{\textbf{C$_{s}$}} \\
\hline

O$(^3\mathrm{P})$ + C$_2(\mathrm{X}\,^1\Sigma_\mathrm{g}^+)$
& 3
& $\Sigma^{-}$, $\Pi$
& $1\,A'$,\, $1\,A''$,\, $2\,A''$  \\
\hline

O$(^3\mathrm{P})$ + C$_2(\mathrm{a}\,^3\Pi_\mathrm{u})$
& 1, 3, 5
& $\Sigma^{+}$, $\Sigma^{-}$, $\Pi$, $\Delta$
& $1\,A'$,\, $2 \, A'$,\, $3\,A'$,\, $1\,A''$,\, $2\,A''$,\, $3\,A''$ \\
\hline

C$(^3\mathrm{P})$ + CO$(\mathrm{X}\,^1\Sigma^+)$
& 3
& $\Sigma^{-}$, $\Pi$
& $1\,A'$,\, $1\,A''$,\, $2\,A''$  \\
\hline

\end{tabular}
\endgroup
\caption{Electronic states arising from the interaction of O$(^3\mathrm{P})$ + C$_2(\mathrm{X}\,^1\Sigma_\mathrm{g}^+, \mathrm{a}\,^3\Pi_\mathrm{u})$ (entrance channel) or C$(^3\mathrm{P})$ + CO$(\mathrm{X}\,^1\Sigma^+)$  (output channel) together with their correlation in C$_{\infty v}$ and C$_{s}$ point groups.}
\label{tabcorrdiag}
\end{table*}

The electronic structure calculations used to construct the PESs were carried out with the \texttt{MOLPRO} 2020 package\cite{molpro2020} in C$_s$ symmetry. The basis set used for both oxygen and carbon atoms is the correlation-consistent polarized valence triple zeta basis sets of Dunning and co-workers\cite{dunning1,dunning2} augmented with diffuse functions for C and O atoms, i.e., aug-cc-pVTZ. The multireference space is constructed using the State-Average Complete Active Space Self-Consistent Field (SA-CASSCF) approach, wherein the valence electrons of the three atoms are distributed across all valence orbitals. This yields an active space of 14 electrons in 12 orbitals (14,12). The 1$s$ orbitals for the three atoms (3$a$') were maintained doubly occupied but previously optimized. The SA-CASSCF calculations were performed including three $^1A'$, three $^1A''$, four $^3A'$, and five $^3A''$ states in the averaging procedure, following preliminary tests that identified this combination as optimal to ensure smooth potential energy surfaces along the reaction coordinates. This choice provided a balanced description of all relevant electronic configurations and reproduced the correct asymptotic degeneracy for both O$(^3\mathrm{P})$ + C$_2(\mathrm{X}\,^1\Sigma_\mathrm{g}^+)$/C$_2(\mathrm{a}\,^3\Pi_\mathrm{u})$ reactant and C$(^3\mathrm{P})$ + CO$(\mathrm{X}\,^1\Sigma^+)$ product channels. Additionally, the calculated $1\,^3A'$ and $2\,^3A''$ states were also found to be degenerate for collinear geometries (within a few meV), in agreement with the expected behavior of $\Pi$ states, confirming the consistency of the active space and the symmetry adaptation of the SA-CASSCF wave functions.
The resulting SA-CASSCF orbitals were subsequently employed as reference functions for the internally contracted multireference configuration interaction (icMRCI) calculations, including single and double excitations\cite{mrci} and the Davidson correction (+Q)\cite{davidson}. Given the highly multiconfigurational nature of this system, the number of configuration state functions (CSFs) required is exceedingly large, making the MRCI calculations computationally prohibitive. Therefore, the reference space is restricted to configurations whose weights exceed a predefined threshold. This threshold was set to 0.005, resulting in energy differences of less than 1\%.	

The  calculated \textit{ab initio} energies (icMRCI+Q) for the electronic states $1\,^3A'$, $1\,^3A''$ and $2\,^3A''$ were individually fitted using the \texttt{GFIT3C} method, as detailed in Refs. \cite{gilles_jcp_2005,fit1,fit2,fit3}. This procedure employs a many-body expansion to represent the global $V_{ABC}$ PES as
\begin{equation}
    V_{ABC} = \sum_{A}^{3}{V_{A}^{(1)}} + \sum_{AB}^{3}{V_{AB}^{(2)}(r_{AB})} + \sum_{n=1}^{N_{\rm 3B}}V_{ABC}^{(3),n}(r_{AB},r_{AC},r_{BC})
    \label{eq:qgfit3c}
\end{equation}
where $V_A^{(1)}$ denotes the energy of atom $A$ in its ground electronic state. The one-body terms are defined such that the energy of the three separated atoms, $E_{\mathrm{O}(^3\mathrm{P})}+2E_{\mathrm{C}(^3\mathrm{P})}$, corresponds to the zero of energy. $V_{AB}^{(2)}(r_{AB})$ represents the diatomic interaction, which vanishes as $r_{AB}\to\infty$ and diverges as $r_{AB}\to0$, whereas $V_{ABC}^{(3),n}(r_{AB},r_{AC},r_{BC})$ denotes the $n$th three-body contribution, which vanishes at all dissociation limits and whenever any internuclear distance approaches zero.   $N_{\mathrm{3B}}$ is the total number of three-body contributions included in the fit. The two- and three-body components, described in the following sections, were adapted from the original \texttt{GFIT3C} formalism to properly describe the complex electronic structure and near-degenerate states of the O + C$_2$ system.

\subsubsection{The two-body potentials}

Each two-body contribution, $V_{AB}^{(2)}(r_{AB})$, describes the interaction between two atoms by characterizing each diatomic fragment (AB = OC, OC, CC). Following the original \texttt{GFIT3C} formalism, this term is expressed as the sum of short- and long-range components,
\begin{equation}
V_{AB}^{(2)}(r_{AB}) =    V_{short}(r_{AB})  +   V_{long}(r_{AB})
\end{equation}
where the short-range potential accounts for the strong repulsive interaction at small internuclear separations and is modeled as a Coulomb potential,
\begin{equation}
  V_{short}(r_{AB})  = \frac{c_0 \cdot e^{-\alpha r_{AB}}}{r_{AB}} \quad\quad (c_0, \alpha > 0)
\end{equation}
tending to infinite as $r_{AB} \rightarrow 0$. The long-range term,
\begin{equation}
  V_{long}(r_{AB})  = \sum_{i}^{L}{c_i \rho_{AB}^{i}} \quad\quad (\beta_{AB} > 0)
\end{equation}
is expressed using modified Rydberg functions, $\rho_{AB}(r_{AB}) = r_{AB}\, e^{-\beta_{AB}^{(2)} r_{AB}} $, which vary with the internuclear distances, approaching zero as the corresponding distance approaches zero or infinity. The $c_{i}\{i=0,...,L\}$ are linear fitting parameters, and $\alpha$ and $\beta_{AB}^{(2)}$ are non-linear fitting parameters.

Accurate representation of the diatomic potentials is particularly important because the three-body term is obtained from the difference between the \textit{ab initio} energies and the sum of the fitted two-body contributions. Errors in the diatomic fits would therefore translate directly into spurious oscillations in the three-body surface. To avoid this, the two-body potentials have to be fitted with a root mean square (rms) error at least twice smaller than the target accuracy of the three-body term, as commonly taken into account in previous \texttt{GFIT3C}-based fits (see, for example, Refs.\cite{zanchet_pccp_2009,susana_chemphyslet_2007,jav_pccp_2023,zanchet_jcp_2019}). In the present system, the diatomic contributions correspond to the OC and CC fragments. The  C$_2$ and CO diatomic potentials are common  for the three triplet surfaces ($1\,^3A'$, $1\,^3A''$, and $2\,^3A''$), as they correlate adiabatically with the same entrance and product channels (see Table \ref{tabcorrdiag}). While the CO fitted potential was directly obtained from the \textit{ab initio} data, the CC diatomic required special treatment due to the near degeneracy between its ground singlet ($\mathrm{X}\,^1\Sigma_g^+$) and first excited triplet ($\mathrm{a}\,^3\Pi_u$) states. To properly account for the interaction between these two electronic terms, we implemented a two-state diabatic-to-adiabatic\cite{beaer_chphlett_1975,beaer_molphys_1980} transformation within the \texttt{GFIT3C} scheme. In this approach, the diabatic singlet and triplet states of C$_2$ are coupled through an off-diagonal spin--orbit interaction term, $V^{^1\Sigma - ^3\Pi}$, leading to a $2\times2$ Hamiltonian matrix:
\begin{equation}
H  =
\begin{pmatrix}
E^{\mathrm{\,^1\Sigma_g^+}}   & V^{^1\Sigma - ^3\Pi} \\
V^{^1\Sigma - ^3\Pi}  & E^{\,^3\Pi_\mathrm{u}}
\end{pmatrix}
\end{equation}
where $E^{\mathrm{\,^1\Sigma_g^+}} $ and $E^{\,^3\Pi_\mathrm{u}} $ are the fitted diabatic potentials for the $\mathrm{X}\,^1\Sigma_g^+$ and $\mathrm{a}\,^3\Pi_\mathrm{u}$ states, respectively. The diagonalization of $H$ yields to two adiabatic energy curves,

 \begin{equation}
 \begin{aligned}
 E_{\pm} &= \frac{1}{2}\! \Bigg(
      E^{\mathrm{\,^1\Sigma_g^+}}  + E^{\,^3\Pi_\mathrm{u}}  \\
      &\quad \pm
      \sqrt{\big(  E^{\mathrm{\,^1\Sigma_g^+}}  - E^{\,^3\Pi_\mathrm{u}}  \big)^2
      + 4 \big(V^{^1\Sigma - ^3\Pi}\big)^2 }
\Bigg)
 \end{aligned}
 \end{equation}
which describe the avoided crossing between the singlet and triplet manifolds. In the present work, $V^{^1\Sigma - ^3\Pi}$ was set to zero, corresponding to the spin-orbit-free limit where the two curves cross diabatically. The lower adiabatic curve, $E_{-}$, was adopted as the effective CC two-body term in the many-body expansion, ensuring a continuous potential energy function across the crossing region while preserving the correct spin character on each side of the intersection. Although this approach introduces a cusp in the derivative of $E_{-}$ at the crossing, the global analytical fit preserves its physical character, so for the subsequent dynamical calculations this region will be smoothed by considering a small effective singlet-triplet coupling  term $V^{^1\Sigma - ^3\Pi}$ to ensure differentiability of the global PES.

\begin{figure}[ht!]
  \centering
  \includegraphics[width=1.0\linewidth]{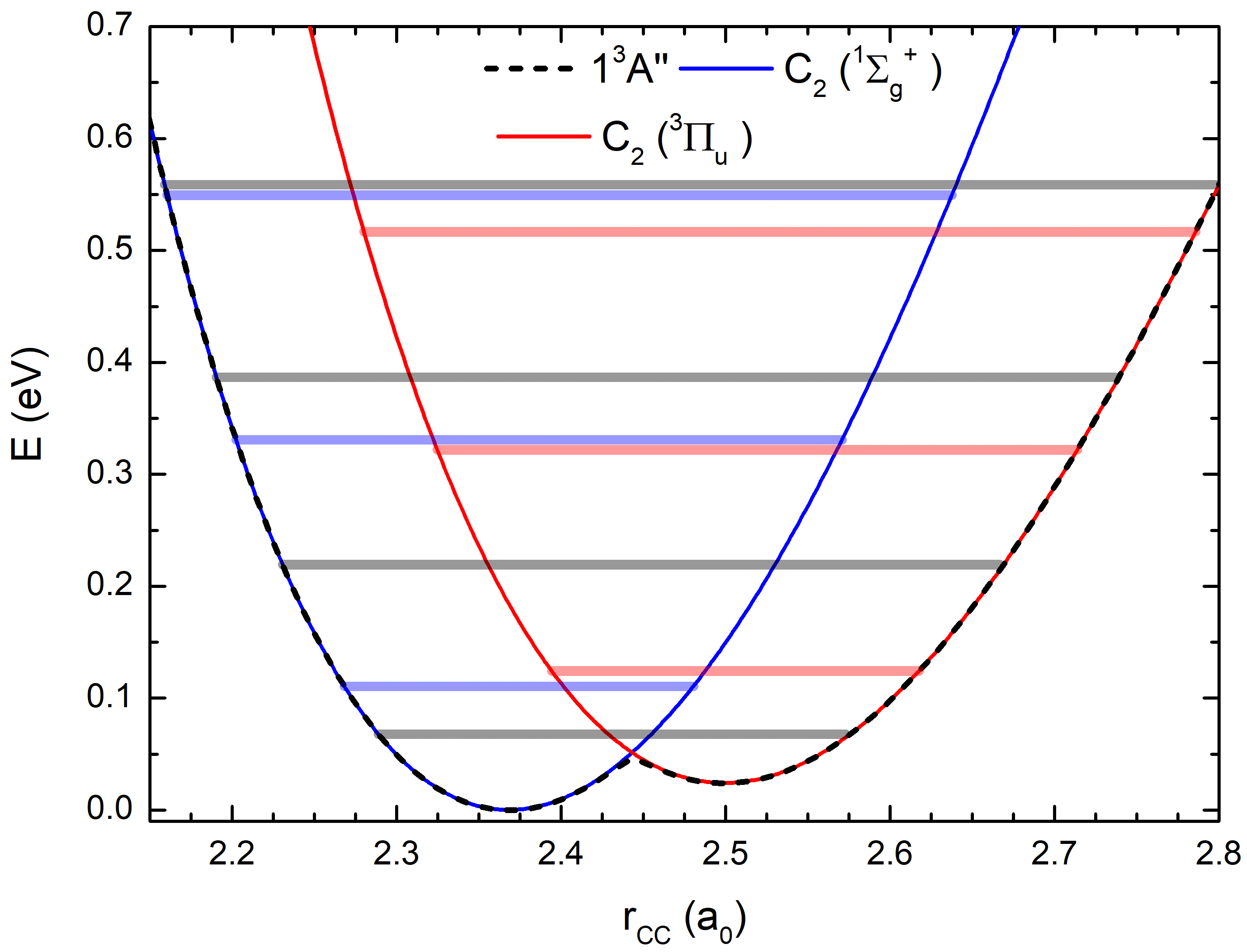}
  \caption{Potential energy curves for the asymptotic C$_2$($^1\Sigma_{g}^{+}$), C$_2$($^3\Pi_u$) molecules (solid blue and red lines) along with the entrance channel of the $1 ^3A''$ PES (dashed black line). Energies are referred to the equilibrium geometry of the C$_2$($^1\Sigma_{\mathrm{g}}^{+}$) state. Vibrational energy levels are represented as horizontal transparent lines using the same color scheme.   }
  \label{fig4}
\end{figure}

This behavior is illustrated in Fig.\ref{fig4}, which shows the fitted singlet and triplet C$_2$ potentials together with the $1\,^3A''$ PES at the asymptotic region (same results are obtained for the other two PESs). As can be seen, the long-range limit of the fitted $1\,^3A''$ PES (with the O atom placed far from the C$_2$ fragment) accurately reproduces the characteristic double-well structure arising from the coupling between the $\mathrm{X}\,^1\Sigma_\mathrm{g}^+$ and $\mathrm{a}\,^3\Pi_u$ states of C$_2$, confirming the  consistency of the diatomic and triatomic fittings.

\subsubsection{The three-body potentials}

Each three-body contribution is represented by an expansion of the form
\begin{equation}
    V_{ABC}^{(3),n}(r_{AB},r_{AC},r_{BC})
    =
    \sum_{ijk}^{K}
    d_{ijk}^{(n)}
    \left[\rho_{AB}^{(n)}\right]^i
    \left[\rho_{AC}^{(n)}\right]^j
    \left[\rho_{BC}^{(n)}\right]^k
    \label{eq:3body}
\end{equation}
where $d_{ijk}^{(n)}$ are the linear coefficients of the $n$th three-body contribution, and $\rho_{XY}^{(n)}$ are the corresponding Rydberg functions of the interatomic distances $r_{XY}$, with independent nonlinear parameters for each contribution. Each expansion is truncated such that $i+j+k\leq K$, where $K$ controls the order of the expansion. The \texttt{GFIT3C} fitting procedure determines the linear coefficients $d_{ijk}^{(n)}$ and the nonlinear parameters of the Rydberg functions by fitting the \textit{ab initio} energies after subtracting the one- and two-body contributions. For a symmetric ABB system such as O$+$C$_2$, permutation symmetry with respect to the two equivalent carbon atoms is imposed on the linear coefficients, reducing the number of independent parameters required in the fit. Further details of the standard fitting procedure can be found in Refs. \cite{fit1,fit2,fit3}.

In the standard \texttt{GFIT3C} formalism, the three-body interaction is represented using a single three-body term, i.e., $N_{\mathrm{3B}}=1$, as done here for the $1\,^3A''$ and $2\,^3A''$ PESs. However, for the $1\,^3A'$ state, a slight modification was required to avoid artificial oscillations in the fitted PES. In this case, the large exothermicity of the reaction and the depth of the potential wells led to small spurious barriers of about 4 meV near the entrance channel. To increase the flexibility of the fit, an additional three-body term of the same analytical form as Eq. (\ref{eq:3body}) was introduced, i.e., $N_{\mathrm{3B}}=2$, but with independent nonlinear parameters $\beta_{\mathrm{CC}}$ and $\beta_{\mathrm{CO}}$. This extension effectively improved the smoothness of the potential around the long-range region without compromising the description of the short-range interaction or the overall accuracy of the global fit.

Similar implementations including multiple three-body terms have also been employed in other systems,  where the use of additional three-body expansions was essential to accurately reproduce shallow minima (CaHCl\cite{gilles_jcp_2005}) or subtle features in the interaction region, such as van der Waals wells (LiHF\cite{alfredo_jcp_2003_bis,cris_chemphyslett_2002}, H$_3^+$\cite{alfredo_pccp_2021,alfredo_jcp_2000,cris_jcp_2001} or H$_4^+$\cite{cris_jcp_2013}).

\subsubsection{Long-range description}

Within the \texttt{GFIT3C} formalism, the long-range behavior is described implicitly through the exponential decay of the Rydberg functions, without introducing explicit analytical dispersion terms. This approach ensures a smooth decay of the potential at large distances but does not provide a physically rigorous asymptotic description beyond the range of the \textit{ab initio} data.

For the present study particular attention was given to the entrance channel O + C$_2$, where the open-shell character of the O($^3\mathrm{P}$) atom and the near degenerate singlet-triplet manifold of C$_2$ make the description of the entrance channel particularly critical. The relevant long-range contributions correspond to $R^{-5}$ (electrostatic quadrupole--quadrupole), $R^{-6}$ (dipole-induced/dipole-induced dispersion), where $R$ is the distance between the oxygen atom and the center of mass of the C$_2$ molecule. These long-range terms define the correct asymptotic behavior that must be recovered in the fitted potential. The  $R^{-8}$ (quadrupole--induced-dipole induction) is comparatively weaker and represents a minor high-order correction, reason why it is not included in the description of the long-range. Based on former discussion, the analytical potential includes the next long-range correction term, using reactant Jacobi coordinates:
\begin{equation}
V_\mathrm{LR}(R,\gamma) =
\frac{\alpha_0 + \alpha_1 \sin(\gamma)}{R^{5}}
+ \frac{\beta_0 + \beta_1 \sin(\gamma)}{R^{6}}
\label{eq:longrange}
\end{equation}
where linear coefficients $\alpha_i$ and  $\beta_i$ are optimised/fitted as a function of $R$ and $\gamma$, keeping r$_{CC}$ fixed in the equilibrium internuclear distance of C$_2$ molecule. This long-range function was independently determined for each of the $1\,A'$, $1\,A''$ and $2\,A''$ PESs.

The final potential is therefore defined as
\begin{equation}
\begin{aligned}
V (r_{\mathrm{CC'}},r_{\mathrm{CO}}, r_{\mathrm{C'O}})
&= V_{ABC}(r_{\mathrm{CC'}},r_{\mathrm{CO}}, r_{\mathrm{C'O}})
+ V_{\mathrm{LR}}\!\big(R_{\mathrm{eff}}(R),\,\gamma\big)
\end{aligned}
\end{equation}
where $ V_{ABC}(r_{\mathrm{CC'}},r_{\mathrm{CO}}, r_{\mathrm{C'O}})$ is given by Eq. (\ref{eq:qgfit3c}), and $V_{\mathrm{LR}}\big(R_{\mathrm{eff}}(R),\gamma\big)$ is the analytical long-range contribution. In  $V_{\mathrm{LR}}$, the atom-diatom distance $R$ is replaced by the effective distance
\begin{equation}R_{\mathrm{eff}}(R)=R+L\,e^{-\epsilon(R-R_0)}
\label{eq}\end{equation}

For all three PESs, we used $L = 20$ $a_0$, $\epsilon = 0.6$ $a_0^{-1}$ and $R_{0} =1.8$ $a_0$. Thsese values were determined empirically through successive tests, optimizing the smooth connection between $V_{ABC}(r_{\mathrm{CC'}},r_{\mathrm{CO}}, r_{\mathrm{C'O}})$ and $V_{\mathrm{LR}}\big(R_{\mathrm{eff}}(R),\,\gamma\big)$. This effective-distance procedure, previously used for other triatomic systems such as H$_3^+$ \cite{alfredo_pccp_2021}, avoids dampings functions,  which can in some cases introduce discontinuities in the derivatives. The resulting global PES smoothly recovers the correct asymptotic limit while preserving the accuracy of the fitted data in the interaction region.

\subsection{Dynamical calculations}

We carried out adiabatic quasiclassical trajectory (QCT) calculations on the three computed PESs using the procedure described in Refs.~\cite{qct1,qct2}. The excitation functions were calculated by running batches of $10^6$ trajectories for the C$_2(^1\Sigma_{\mathrm{g}}^{+})$ and C$_2(^3\Pi_{\mathrm{u}})$ states at different initial rovibrational levels ($v=0, j=0$ and $v=3, j=0, 50$). Furthermore, to highlight the  differences in reactivity between the two C$_2$ states, we also computed the excitation functions starting the trajectories from the bottom of the C$_2$ potential energy curves, which corresponds to an artificial $v=-0.5$ state. Trajectories were initiated at an atom--diatom distance of 10~\AA{} with an integration time step of $3 \times 10^{-17}$~s, guaranteeing total energy conservation to better than one part in $10^5$. For low collision energies ($E_{\mathrm{col}}$) between 0.001 and 0.1 eV, additional batches of $10^6$ trajectories were propagated using the same integration step but a larger initial separation of 20~\AA{}. Finally, cumulative reaction probabilities for total angular momentum $J=0$ were calculated following the methodology detailed in Refs.~\cite{Aoiz:JCP06,Jambrina:JCP09}.

The selection of the initial conditions of the trajectories deserves further comments. As it can be appreciated in Fig. \ref{fig4} the C$_2$ ($^1\Sigma_{g}^{+}$) and C$_2$ ($^3\Pi_{u}$) states are not well isolated in the global calculated PES, which depicts a double well structure similar to that found, for example, for the umbrella motion of ammonia. To initialize the trajectories, we computed the turning points and vibrational energies using the isolated C$_2(^1\Sigma_{\mathrm{g}}^{+})$ and C$_2(^3\Pi_{\mathrm{u}})$ diatomic potentials, and subsequently propagated the trajectories on the global PESs. This ``double-well'' structure also affects how we specified the initial internuclear separation and vibrational phase of the QCT calculations.
A standard and convenient method for setting these variables is to select the initial internuclear distance of the diatomic molecule randomly between its inner and outer turning points, while adjusting the initial atom--diatom separation for each trajectory. This ensures a correct sampling of the vibrational phase when the incoming atom reaches the nominal starting distance.\cite{truhlar79} The advantage of this approach is that it avoids introducing a radial component to the diatomic rotational angular momentum. In our case, however, instead of sampling randomly between the inner and outer turning points, we strictly select the inner turning point for the C$_2(^1\Sigma_{\mathrm{g}}^{+})$ state and the outer turning point for the C$_2(^3\Pi_{\mathrm{u}})$ state. At these specific geometries, the isolated diatomic curves of the respective states perfectly match the global PESs.

\section{Results}\label{sec:results}
\subsection{Potential Energy Surface}

As a starting point, Fig.~\ref{fig4} displays the diatomic potential energy curves for the C$_2(^1\Sigma_{\mathrm{g}}^{+})$ (blue line) and C$_2(^3\Pi_{\mathrm{u}})$ (red line) states. A prominent feature of the C$_2$ molecule is the near-degeneracy of these two lowest-lying electronic states, with their crossing point lying below the zero-point energy of both states. The lowest vibrational levels are indicated in Fig.~\ref{fig4} by horizontal lines corresponding to the color of each state. The figure also illustrates the asymptotic potential energy curve of the $1\,^3A''$ global PES, which smoothly connects these states; it perfectly matches the C$_2(^1\Sigma_{\mathrm{g}}^{+})$ potential at internuclear distances ($r_{\mathrm{CC}}$) below 2.43~$a_0$ and follows the C$_2(^3\Pi_{\mathrm{u}})$ potential at larger $r_{\mathrm{CC}}$ values. Due to the lower vibrational constant of C$_2(^3\Pi_{\mathrm{u}})$ , while the $v=0$ level of the C$_2(^1\Sigma_{\mathrm{g}}^{+})$ state is lower in energy than that of the C$_2(^3\Pi_{\mathrm{u}})$ state, this energy ordering is reversed for vibrationally excited levels.

The accuracy of the analytical PESs was evaluated by comparing the fitted energies with the corresponding \textit{ab initio} values over different regions of the configuration space. Table~\ref{tab:diatoms} reports the equilibrium bond distance ($r_e$), the exothermicity of the reaction ($\Delta D_e$), and the adiabatic electronic excitation energy from the ground state ($T_e$, when applicable) computed at the icMRCI+Q level, comparing them with both the analytical diatomic fits used in the many-body expansion and available experimental data. The values in Table~\ref{tab:diatoms} show that the fit accurately reproduces the icMRCI+Q results for $r_e$, $\Delta D_e$, and $T_e$. The icMRCI+Q and fitted values are also in good agreement with the experimental values for $r_e$ and $\Delta D_e$, whereas both underestimate $T_e$. Table~\ref{tab:triatomic} reports the root-mean-square (rms) errors for several energy ranges across the three considered electronic states ($1\,^3A''$, $2\,^3A''$, and $1\,^3A'$). The overall rms errors of the fitted PESs are 0.022, 0.026, and 0.028 eV for the $1\,^3A''$, $2\,^3A''$, and $1\,^3A'$ states, respectively.

\begin{figure}
  \centering
  \includegraphics[width=1.0\linewidth]{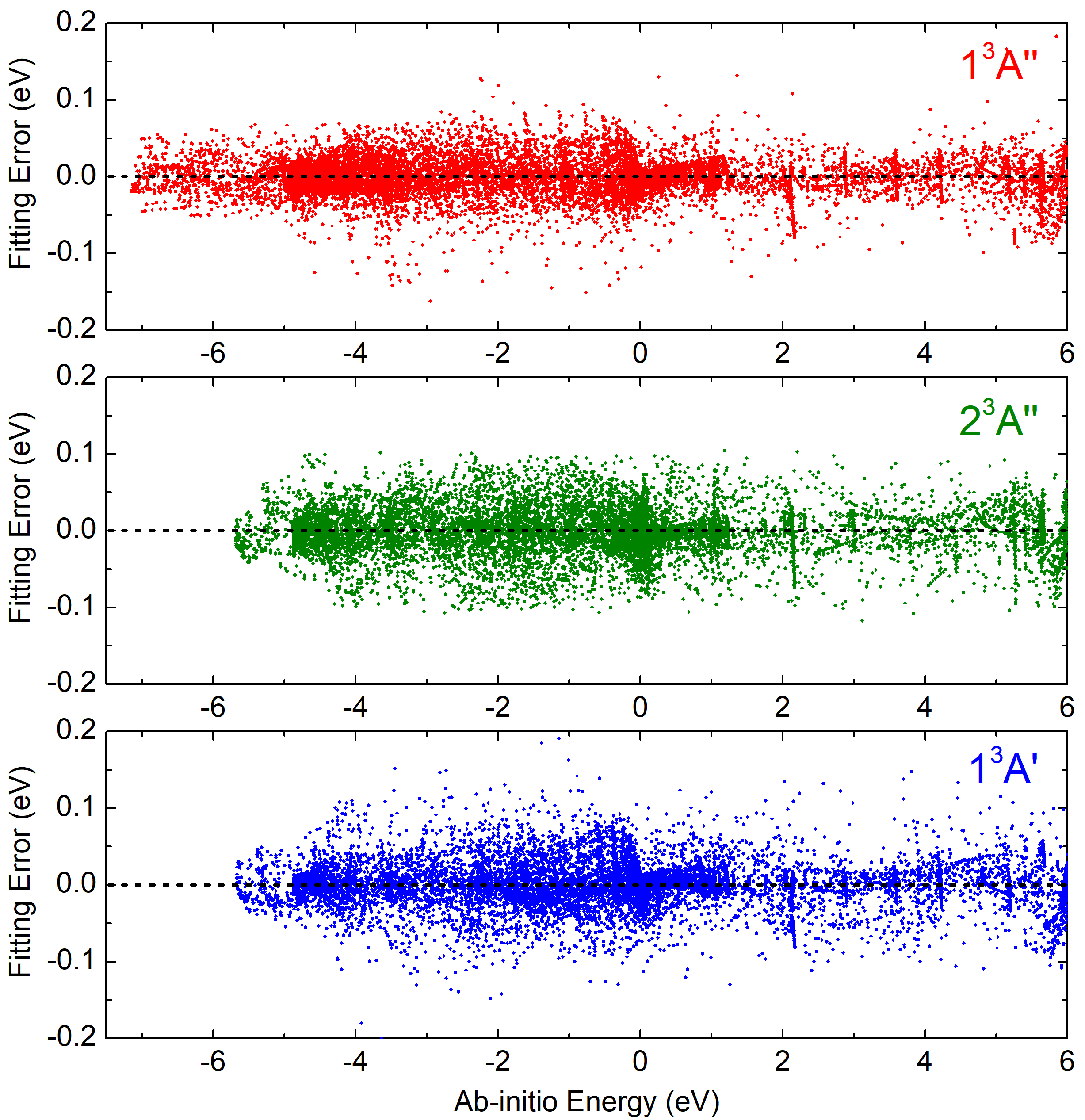}
  \caption{Distribution of the fitting errors ($E_{\rm fit} - E_{\rm ab initio}$ ) as a function of the \textit{ab initio} energy for the 3 PESs calculated (top panel: ${1\, ^3A''}$, middle panel: ${2\, ^3A''}$, bottom panel: ${1\, ^3A'}$)   }
  \label{fig:errors}
\end{figure}

The fitted PESs cover energies from the global minimum of each surface up to values above the three-atom dissociation limit ($\approx 6$~eV, see Fig.~\ref{fig:diagram}), which represents an unusually large domain for a global fit. A detailed examination of Table~\ref{tab:triatomic} shows that the smallest deviations are obtained in the energy region around the global minima. The rms errors increase in the high-energy interval ($E > 0.1$~eV) and especially at short internuclear distances, irrelevant to the dynamics, where the repulsive wall is steep and fewer \textit{ab initio} points are available. The smaller deviations obtained for the $1\,^3A''$ PES are consistent with the relative topological complexity of the three electronic states. The $1\,^3A''$ PES presents a comparatively simple structure  dominated by a single deep well, and it exhibits a rather regular behavior across the relevant geometries, which facilitates its analytical representation. In contrast, the $2\,^3A''$ and $1\,^3A'$ states present a more complex behavior, featuring competing wells, saddle-point regions, and a strong geometric dependence as a consequence of avoided crossing with higher-energy PESs, making their analytical description more demanding.
A complementary representation of the fitting quality for the three PESs is provided in Fig.~\ref{fig:errors}, which illustrates the distribution of the fitting errors as a function of the \textit{ab initio} energies. As observed, a slightly larger dispersion is obtained for the $1\,^3A'$ PES whereas the errors for all three PESs remain narrowly distributed around zero in the energetically relevant regions.

\begin{table}[ht!]
\centering
\caption{Spectroscopic properties of C$_2(\mathrm{X}\,^1\Sigma_\mathrm{g}^+)$, C$_2(\mathrm{a}\,^3\Pi_\mathrm{u})$, and CO$(\mathrm{X}\,^1\Sigma^+)$. A comparison between the fitted, calculated (icMRCI+Q) and experimental data. The exothermicity of the reaction is calculated as: $\Delta D_e = D_e[\mathrm{C}_2(\mathrm{X}\,^1\Sigma_{\mathrm{g}}^+)] - D_e[\mathrm{CO}(\mathrm{X}\,^1\Sigma^+)]$, and $T_e$ corresponds to the energy difference between C$_2(^1\Sigma_\mathrm{g}^+)$ and C$_2(^3\Pi_\mathrm{u})$ states. Bond distances are given in $a_0$ and energies in eV.}
\renewcommand{\arraystretch}{1.2}
\begin{tabular}{l  l | c | c | c |}
\toprule
\textbf{Diatom} &  & \textbf{C$_2(\mathrm{X}\,^1\Sigma_\mathrm{g}^+)$} & \textbf{C$_2(\mathrm{a}\,^3\Pi_\mathrm{u})$} & \textbf{CO$(\mathrm{X}\,^1\Sigma^+)$} \\
\midrule
\multirow{3}{*}{$r_e$}
    & icMRCI+Q  $\quad$      &  2.365    &  2.500  & 2.15 \\
    & Fit                   & 2.3675   &  2.5009 &  2.1487 \\
    & Exp.      & 2.3495$^a$ & 2.479$^b$ & 2.1321$^d$ \\
\midrule
\multirow{3}{*}{$\Delta D_e$/ $T_e$}
    & icMRCI+Q    & $-$ & 0.028 & $-4.867$ \\
    & Fit & $-$ & 0.024 & $-4.870$ \\
    & Exp.      & $-$ & 0.089$^{b,c}$ & $-4.80^{a,d}$\\
\bottomrule
\end{tabular}
\label{tab:diatoms}
\vspace{0.3em} %
\parbox{0.95\linewidth}{\raggedright \footnotesize
$^a$ Ref. \cite{boschen_theochemacc_2014} \,  
$^b$ Ref. \cite{kirby_jchemphys_1979} \,  
$^c$ Ref. \cite{veset_canjphys_1975} \,  
$^d$ Ref. \cite{lu_commtheophys_2013} 

}
\end{table}

 \begin{table}[ht!]
    \centering
    \caption{Root mean-square (rms) deviations between the \textit{ab initio} and fitted energies for several energy ranges of the $1\,^3A''$, $2\,^3A''$, and $1\,^3A'$ states of OC$_2$. Energy ranges and rms values are given in eV, and $N$ is the number of \textit{ab initio} points.}
    \setlength{\tabcolsep}{4pt}
    \begin{tabular*}{\linewidth}{@{\extracolsep{\fill}}lcc}
    \toprule
    \toprule
    \vspace{0.05cm}
    \\[-0.10cm]
    \multicolumn{3}{c}{\textbf{${1\, ^3A''}$}} \\
        \vspace{-0.2cm}
    \\[-0.10cm]
    \midrule
     & N & rms \\
 \textbf{All data points:}    & 18684  & 0.022      \\
    \textbf{Energy range:} &  &  \\
       \( \,\,\, E > 0.10\)           & 5144   & 0.024      \\
    \(\,\,\, -2.00 < E < 0.10\)     &  6041   & 0.020     \\
    \(\,\,\, -5.50 < E < -2.00\) & 6845   & 0.023      \\
        \(\,\,\, E < -5.50\) & 654   & 0.022     \\
    \midrule
    \midrule
    \vspace{0.05cm}
    \\[-0.02cm]
    \multicolumn{3}{c}{\textbf{\(2\, ^3A''\)}} \\
       \vspace{-0.2cm}
    \\[-0.10cm]
    \midrule
     & N & rms  \\
 \textbf{All data points:}    & 17652  & 0.026     \\
    \textbf{Energy range:} &  &   \\
       \( E > 0.10\)           & 5084   & 0.026     \\
    \(-2.00 < E < 0.10\)   & 7022   & 0.025       \\
    \(-4.50 < E < -2.00\) & 3391   & 0.033       \\
        \(E < -4.50\) & 2155   & 0.019      \\
    \midrule
    \midrule
    \vspace{0.05cm}
    \\[-0.02cm]
    \multicolumn{3}{c}{\textbf{\(1\, ^3A'\)}} \\
        \vspace{-0.2cm}
    \\[-0.10cm]
    \midrule
      & N & rms  \\
 \textbf{All data points:}    & 16056  & 0.028      \\
    \textbf{Energy range:} &  &  \\
       \( E > 0.10\)           & 5854   & 0.029    \\
    \(-2.00 < E < 0.10\)   & 6336   & 0.026      \\
    \(-4.50 < E < -2.00\) & 2360   & 0.036       \\
        \(E < -4.50\) & 1506   & 0.014      \\
    \bottomrule
    \bottomrule
    \end{tabular*}
\label{tab:triatomic}
\end{table}

 \begin{table}[ht!]
    \centering
    \caption{Comparison of the stationary points on the three PESs of the CCO system with previous studies. The tabulated relative energies are
defined as $E_{\mathrm{i}}-E_{\mathrm{O+CC'}}$ and
$E_{\mathrm{i}}-E_{\mathrm{C'+CO}}$, where $E_{\mathrm{i}}$
denotes the energy of the corresponding stationary point, and
$E_{\mathrm{O+CC'}}$ and $E_{\mathrm{C'+CO}}$ denote the energies of
the reactant and product asymptotes, respectively. For the present
fitted PESs, $E_{\mathrm{O+CC'}}=0$ and
$E_{\mathrm{C'+CO}}=\Delta D_e=-4.870$~eV, as reported in
Table~\ref{tab:diatoms}. Distances are given in $a_0$ and energies in eV.}
    \setlength{\tabcolsep}{4pt}
    \begin{tabular*}{\linewidth}{@{\extracolsep{\fill}}lcccc}
    \toprule
    \toprule
    \vspace{0.05cm}
    \\[-0.10cm]
    \multicolumn{5}{c}{\textbf{${1\, ^3A''}$}} \\
        \vspace{-0.2cm}
    \\[-0.10cm]
    \midrule
        & $r_{\,\rm CC'}$ & $r_{\,\rm CO}$ & $E_{\mathrm{i}}-E_{\mathrm{O+CC'}}$  & $E_{\mathrm{i}}-E_{\mathrm{C'+CO}}$ \\
    \textbf{Global minimum:} &   &  &   &  \\
   This work        & 2.587  & 2.212   & $-$7.16  & $-$2.29 \\
   Huang \textit{et al.}\cite{zang_jchemphys_2015}        & 2.627  & 2.186   & $-$7.10  & $-$2.11 \\
   DFT/B3LYP\cite{schatz_jphyschemc_2012}        & 2.579  & 2.217   & $-$8.56  & $-$2.77 \\
   CCSD(T)/B3LYP\cite{schatz_jphyschemc_2012}        & $-$  & $-$   & $-$7.49  & $-$2.35 \\
   CCSD(T)/aVQZ\cite{pinorios_intjquantumchem_2023}        & 2.585  & 2.175   & $-$7.08  & $-$2.20 \\
   CASPT2\cite{zengin_jchempyhs_1996}        & 2.595  & 2.209   & $-$  & $-$2.40 \\
   Exp.\cite{ohshima_jchemphys_1995}        &  2.590   & 2.183   & $-$ & $-$   \\
   Exp.\cite{choi_jchemphys_1998}        &  $-$   & $-$   & $-$ & $-(2.24 \pm 0.02)$   \\
    \midrule
    \midrule
    \vspace{0.05cm}
    \\[-0.02cm]
    \multicolumn{5}{c}{\textbf{\(2\, ^3A''\)}} \\
            \vspace{-0.2cm}
    \\[-0.10cm]
     \midrule
            & $r_{\,\rm CC'}$ & $r_{\,\rm CO}$ & $E_{\mathrm{i}}-E_{\mathrm{O+CC'}}$  & $E_{\mathrm{i}}-E_{\mathrm{C'+CO}}$  \\
    \textbf{Global minimum:} &  &   &   &   \\
   This work      & 2.387  & 2.256  & $-$5.70  & $-$0.83 \\
         &     &   &  &   \\
    \textbf{Transition state:} &  &  &   &   \\
    This work    & 3.036    & 2.248   & $-$4.32 & $+$0.55   \\
    \midrule
    \midrule
    \vspace{0.05cm}
    \\[-0.02cm]
    \multicolumn{5}{c}{\textbf{\(1\, ^3A'\)}} \\
            \vspace{-0.2cm}
    \\[-0.10cm]
    \midrule
                & $r_{\,\rm CC'}$ & $r_{\,\rm CO}$ & $E_{\mathrm{i}}-E_{\mathrm{O+CC'}}$  & $E_{\mathrm{i}}-E_{\mathrm{C'+CO}}$  \\
    \textbf{Global minimum:} &  &   &   &   \\
    This work      & 2.384  & 2.254   & $-$5.67  & $-$0.80 \\
           &     &   &  &   \\
    \textbf{Transition state:} &  &  &   &   \\
    This work     & 3.222    & 2.288   & $-$3.86 & $+$1.02   \\
    \bottomrule
    \bottomrule
    \end{tabular*}
\label{tab:comparacion}
\end{table}

\begin{figure*}[ht!]
  \centering
  \includegraphics[width=1.0\linewidth]{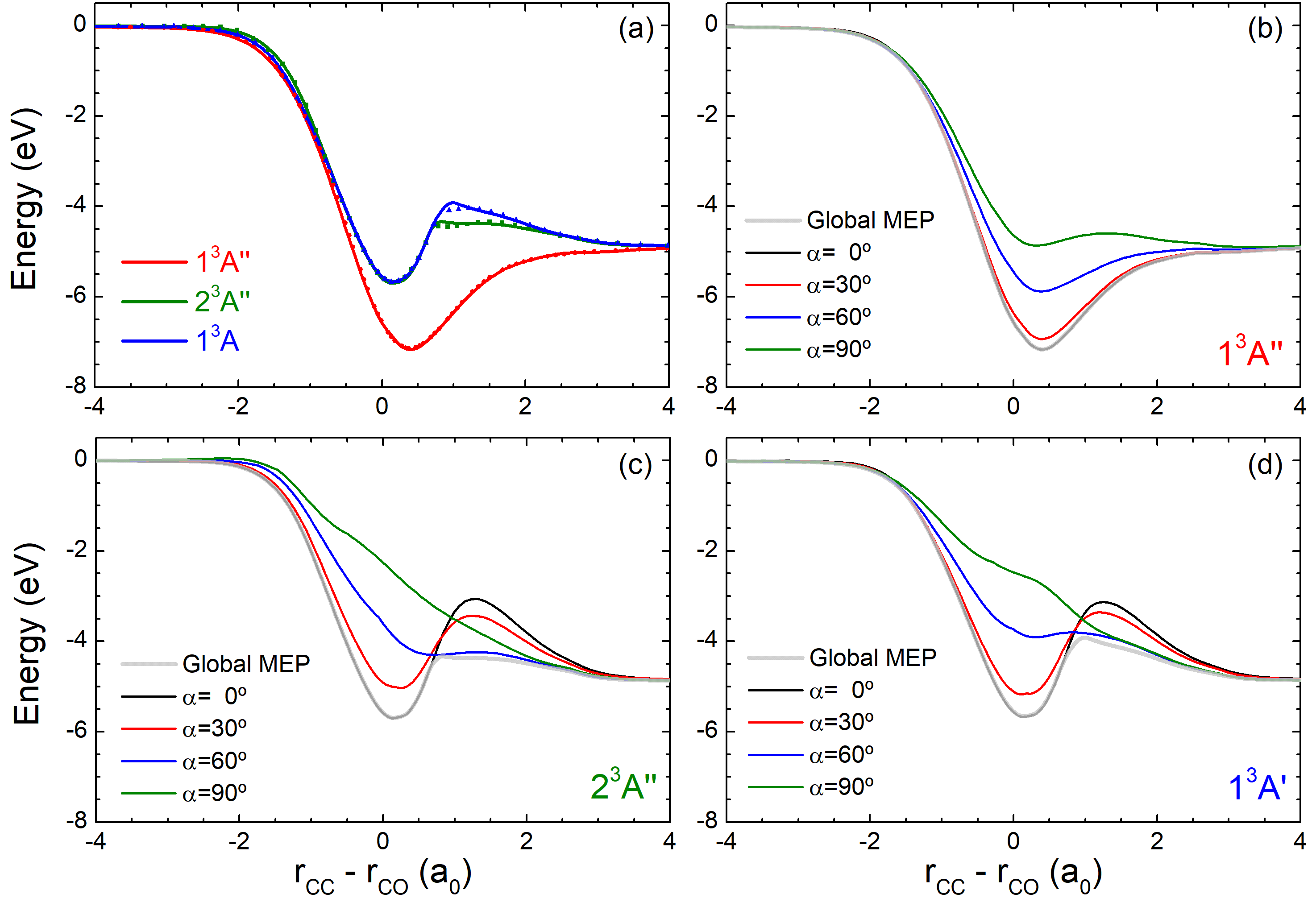}
  \caption{MEPs as a function of the $r_{\rm CC} - r_{\rm CO}$. Panel (a) shows the global MEPs for the three PESs calculated (${1\, ^3A''}$ in red, ${2\, ^3A''}$ in dark green, and  ${1\, ^3A'}$ in blue). Panels (b)$-$(d) show the Global MEPs (in grey) along with the MEPs calculated for different internuclear angles, $\alpha$, which is defined as the C$-$C$-$O angle. Results for ${1\, ^3A''}$ are shown in panel (b), for ${2\, ^3A''}$in panel (c) and for ${1\, ^3A'}$in panel (d). Lines correspond to the analytical PESs, while symbols denote the \textit{ab initio} points. }
  \label{fig:meps}
\end{figure*}

To further illustrate the topological differences discussed above and provide valuable insight into the reaction dynamics, the Minimum Energy Paths (MEPs) of the three triplet PESs are shown in Fig.~\ref{fig:meps}, where the MEPs are calculated as a function of $r_{\mathrm{CC}} - r_{\mathrm{CO}}$, meaning that negative values correspond to geometries close to the reactant asymptote, whereas positive values represent configurations closer to the product channel. Panel (a) of Fig.~\ref{fig:meps} presents the global MEPs for the $1\,^3A''$, $2\,^3A''$, and $1\,^3A'$ states, comparing their analytical fits with the corresponding \textit{ab initio} points and showing a very good agreement.

The $1\,^3A''$ surface exhibits a deep minimum that lies 7.16 eV below the entrance channel (2.29 eV below the product channel) and presents no barrier along the reaction coordinate. This minimum is located at the collinear O--C--C geometry ($\gamma=0^\circ$) with $r_{\mathrm{CC}}=2.587\,a_0$ and $r_{\mathrm{CO}}=2.212\,a_0$. These values are in excellent agreement with both theoretical and experimental results reported in the literature (see Table~\ref{tab:comparacion}). The $2\,^3A''$ and $1\,^3A'$ global MEPs also show no barrier in the entrance channel and a deep minimum for the collinear geometry; however, the potential of the entrance channel is less attractive, and the minima are shallower, lying approximately 5.7 eV and 0.8 eV below the reactant and product asymptotes, respectively. It is worth noting that the $2\,^3A''$ and $1\,^3A'$ PESs are degenerate for collinear geometries, as they constitute the two components of a $\Pi$ state in $C_{\infty v}$ symmetry. In the exit channel, while the $1\,^3A''$ surface shows no barrier, both the $2\,^3A''$ and $1\,^3A'$ PESs display submerged barriers located 0.55 eV and 1.02 eV above the product channel, respectively. Because the geometry at these submerged transition states is no longer collinear, the spatial degeneracy is broken and the two PESs split. The exact geometries of the saddle points and minima are listed in Table~\ref{tab:comparacion}.

Panels (b)-(d) of Fig.~\ref{fig:meps} show the dependence of the MEPs on the C--C--O bending angle, $\alpha$, for each PES. On the $1\,^3A''$ surface, the depth of the well decreases as $\alpha$ increases, and a shallow, submerged barrier emerges in the product channel only for $\alpha > 60^\circ$. The topological changes are more profound on the $2\,^3A''$ and $1\,^3A'$ PESs. While the well depth also decreases with increasing $\alpha$ on these surfaces, they feature a substantial barrier in the exit channel for collinear geometries (approximately 2 eV above the product asymptote). The height of this barrier decreases as the molecule bends, an effect that is particularly pronounced on the $2\,^3A''$ PES. Consequently, although the system favors a collinear approach into the well, it strongly prefers to adopt bent configurations in the exit channel.

The exit-channel barrier arises from an avoided crossing with higher-lying triplet states that correlate with the C$(^3\mathrm{P})$ + CO$(^3\Pi)$ asymptote, which lies approximately 1.4 eV above the O$(^3\mathrm{P})$ + C$_2(^1\Sigma_{\mathrm{g}}^{+})$ entrance channel. The coupling to this higher triplet manifold becomes increasingly significant at non-collinear geometries, producing the avoided-crossing pattern observed in panels (c) and (d) of Fig.~\ref{fig:meps}. For collinear approaches, the relevant states remain degenerate and the barrier reaches its maximum; however, bending breaks this spatial degeneracy and causes the barrier on both the $2\,^3A''$ and $1\,^3A'$ surfaces to decrease progressively. These combined features explain why analytically fitting these two PESs is significantly more challenging than fitting the $1\,^3A''$ state (as reflected by the rms  values in Table~\ref{tab:triatomic}). Specifically, the collinear approach constitutes the most problematic region, where the analytical fit underestimates the \textit{ab initio} energies. Nevertheless, because this region lies well above the product channel, it is energetically irrelevant to the reactive dynamics.

\begin{figure*}
  \centering
 \includegraphics[width=1.0\linewidth]{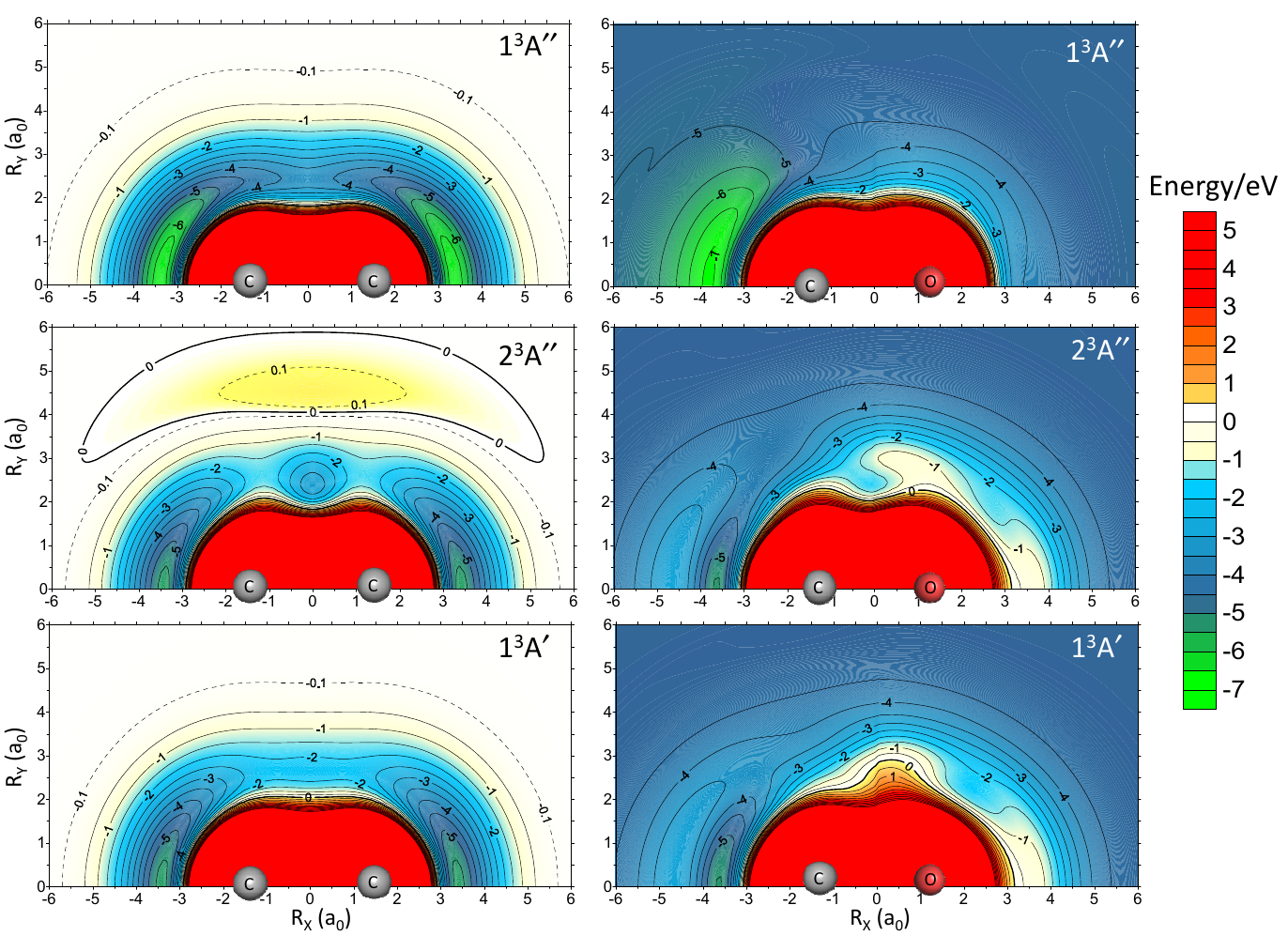}
  \caption{Left panels: Polar plot representation of the entrance channel for C$_2$ at its equilibrium distance. The C$_2$ molecule lies along the x-axis with its center in (0,0). Right panels: Polar plot representation of the exit channel for CO at its equilibrium distance, lying the CO molecule along the x-axis with its center of mass in (0,0). Top panel corresponds to the ${1\, ^3A''}$ PES, middle panel to the ${2\, ^3A''}$ PES , and bottom panel to the ${1\, ^3A'}$ PES. Energies are referred to the reactant asymptote, and the color code is shown in the figure with repulsive contours colored in yellow and red, white is reserved for energies close to that of the reactant channel, blue for negative energies (but above the energy of the product asymptote) and green for energies below the product asymptote.}
  \label{fig:cuts}
\end{figure*}

For barrierless, highly exothermic reactions featuring deep potential wells, such as the dynamics on the $1\,^3A''$, $2\,^3A''$, and $1\,^3A'$ PESs computed here, the exit channel and specially the  entrance channel are the most critical regions of the potential for the dynamics. To highlight the features of this region, the left panels of Fig.~\ref{fig:cuts} display the entrance-channel potential topography as polar $(R_X, R_Y)$ maps. These maps were calculated at a fixed diatom distance of $r_{\mathrm{CC}} = 2.37\,a_0$, which corresponds to the equilibrium bond length of the isolated C$_2 (^1\Sigma_{\mathrm{g}}^{+})$. On the $1\,^3A''$ PES (top-left panel), the approach of the O atom toward either terminal carbon of C$_2$ leads directly to the global minimum at the collinear configuration, with no minima observed for T-shaped geometries, and no barriers to any approach geometry.
The potential on the $2\,^3A''$ PES (middle-left panel) is significantly more anisotropic. For T-shaped approaches, this surface exhibits a barrier of over 0.1 eV with respect to the C$_2$ asymptote. This steric hindrance restricts access to the interaction region and is expected to result in smaller reaction cross sections ($\sigma$) at low collision energies ($E_{\mathrm{col}}$). Additionally, the $2\,^3A''$ PES displays deep wells for collinear geometries alongside a secondary, shallower well for the T-shaped conformation at shorter intermolecular distances. The entrance channel of the $1\,^3A'$ surface (bottom-left panel) is qualitatively similar to that of the $1\,^3A''$ PES, exhibiting no barriers for any approach geometry, although the well depth is substantially shallower, as discussed previously.

The right panels of Fig.~\ref{fig:cuts} display contour cuts of the PESs in the exit channel, calculated at a fixed CO bond length of $r_{\mathrm{CO}} = 2.14\,a_0$, corresponding to the equilibrium geometry of the product diatomic. Regardless of the considered PES, no attractive wells exist for the C--O--C insertion conformation. Furthermore, the submerged transition states in the exit channels of the $2\,^3A''$ and $1\,^3A'$ PESs are clearly visible at $R_X \approx -5\,a_0$ in  collinear limit. In agreement with our previous analysis, these barriers vanish as the geometry bends.

\subsection{Dynamical results}

\begin{figure*}[ht!]
  \centering
  \includegraphics[width=1.0\linewidth]{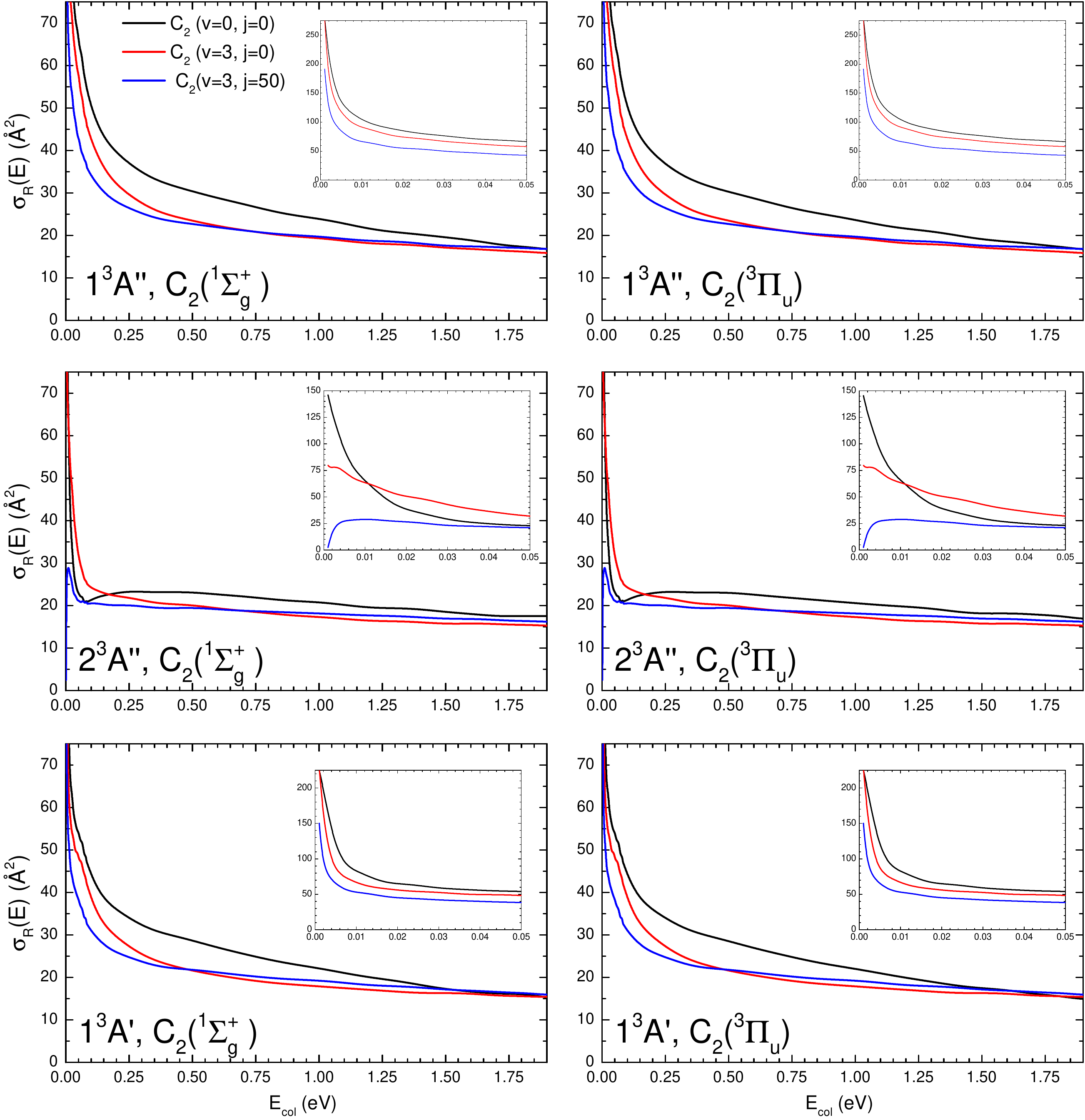}
  \caption{QCT excitation function (cross section as a function of the collision energy) for the O + C$_2$($v$,$j$) reaction for different combinations of ($v$, $j$) on the three PESs.  The left and right panels correspond to C$_2$($^1\Sigma_{\mathrm{g}}^{+},v,j$) and C$_2$($^3\Pi_\mathrm{u},v,j$), respectively. Top, middle and bottom panels show the results obtained on the $1\,^3A''$, $2\,^3A''$ and $1\,^3A'$ PESs, respectively. }
  \label{fig:xsecs-panels}
\end{figure*}

To analyze the dynamical behavior of the O + C$_2$ system, Fig.~\ref{fig:xsecs-panels} presents the excitation functions, $\sigma_{\mathrm{R}}(E_{\mathrm{col}})$, defined as the reactive cross sections as a function of collision energy $E_{\rm col}$. These were obtained from QCT calculations on the three computed PESs: $1\,^3A''$ (top), $2\,^3A''$ (middle), and $1\,^3A'$ (bottom) for different initial rovibrational states. The left and right panels correspond to dynamics initiated from C$_2(^1\Sigma_{\mathrm{g}}^{+})$ and C$_2(^3\Pi_{\mathrm{u}})$, respectively.  With the exception of the calculations on the $2\,^3A''$ PES for highly excited rovibrational states ($v=3, j=50$), which exhibit a small energy threshold of less than 2~meV, all $\sigma_{\mathrm{R}}(E_{\mathrm{col}})$ curves show the expected behavior for exothermic barrierless reactions: no energy threshold and a cross section that decreases monotonically with $E_{\mathrm{col}}$. As expected, $\sigma_{\mathrm{R}}(E_{\mathrm{col}}) (1\,^3A'') > \sigma_{\mathrm{R}}(E_{\mathrm{col}}) (1\,^3A') > \sigma_{\mathrm{R}}(E_{\mathrm{col}}) (2\,^3A'') $, at least for $E_{\rm col}$ < 1.0 eV. For all three PESs, the excitation functions corresponding to a given initial rovibrational state are virtually identical for C$_2$ initially in the $^1\Sigma_{\mathrm{g}}^{+}$ or $^3\Pi_{\mathrm{u}}$ electronic state. Since both C$_2$ electronic states correlate with the same three adiabatic PESs and the $v=0$ level lies above the crossing between their asymptotic potential curves, trajectories initialized from either state evolve over the same classically accessible region of each PES. Consequently, the dynamics is essentially independent of whether C$_2$ is initially in the singlet or triplet state, resulting in nearly identical excitation functions.

These results contrast with those reported in Ref.~\citenum{zang_jchemphys_2015}, where a threshold of approximately 0.08~eV was found for $\sigma_{\mathrm{R}}(E_{\mathrm{col}})$ on the $1\,^3A''$ surface starting from the $(v=0, j=0)$ rovibrational state. This threshold indicated an entrance-channel barrier in their fitted PES. In contrast, our $1\,^3A''$ surface is strictly barrierless, resulting in no threshold for $\sigma_{\mathrm{R}}(E_{\mathrm{col}})$. We suspect that the matching between the C$_2$ asymptotic region and the interaction region in Ref.~\citenum{zang_jchemphys_2015} introduced a small artificial barrier, which likely caused this spurious energy threshold. The absence of an entrance barrier is also consistent with findings for the analogous S$(^3\mathrm{P})$ + C$_2$ reaction on the 2D PES calculated by Rocha \textit{et al.}~\cite{OctavioSC2}.

Finally, on the $1\,^3A''$ and $1\,^3A'$ PESs, C$_2$ vibrational and rotational excitation leads to a decrease in $\sigma_{\mathrm{R}}(E_{\mathrm{col}})$. For a typical insertion reaction, one might expect vibrational excitation to facilitate the insertion of the attacking atom into the diatomic bond. However, in this case, the minimum of the well does not correspond to the insertion of O into the C$_2$ bond, but rather to an O--C--C approach at an $r_{\mathrm{CC}}$ distance shorter than the outer turning point of the $v=0$ state. Hence, vibrational excitation, which increases the probability of the C$_2$ molecule adopting extended internuclear configurations, hinders optimal access to the well, leading to a decrease in the $\sigma_{\mathrm{R}}(E_{\mathrm{col}})$.

\begin{figure}[ht!]
  \centering
  \includegraphics[width=1.0\linewidth]{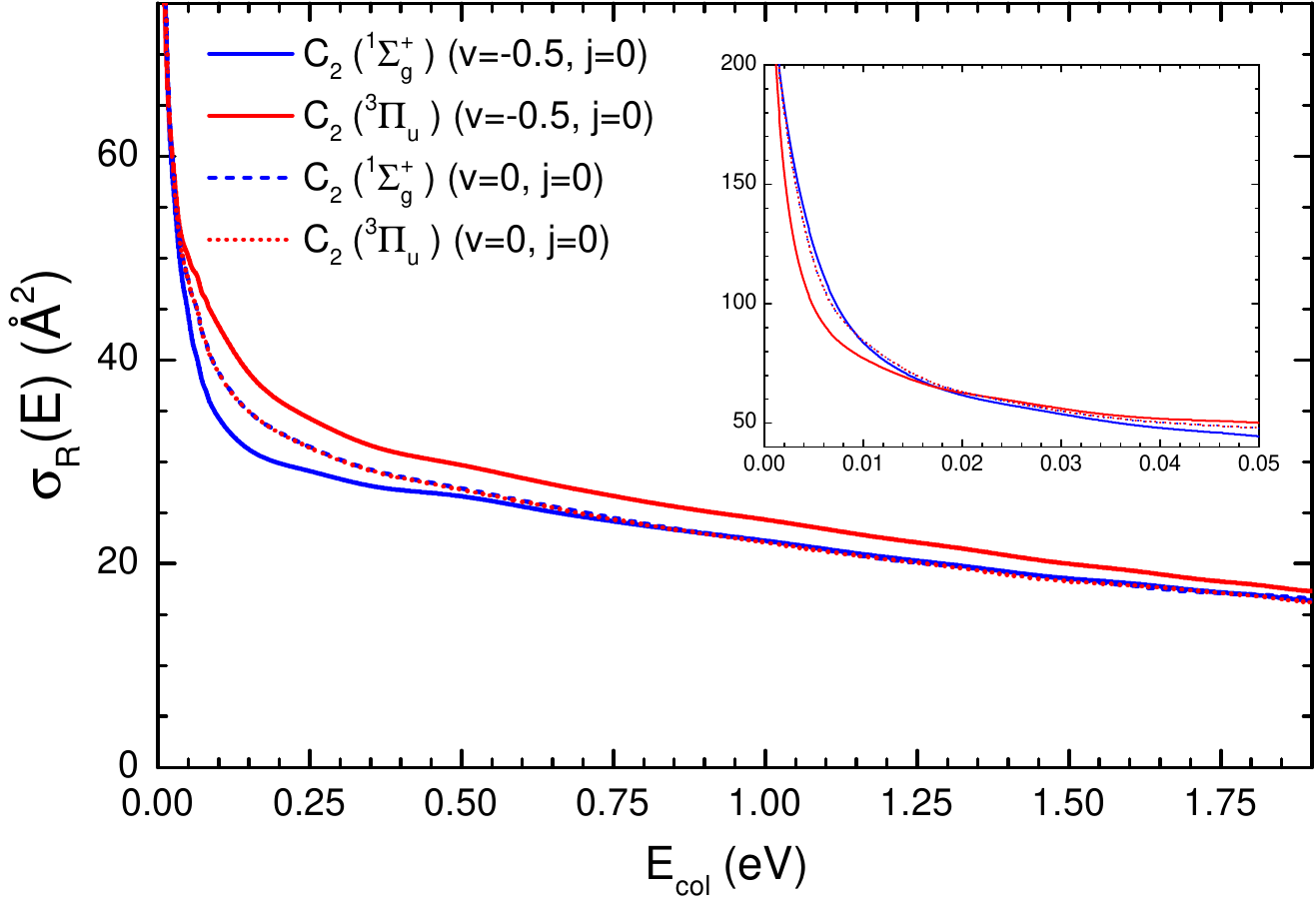}
  \caption{QCT excitation functions averaged over the $1\,^3A''$, $2\,^3A''$ and $1\,^3A'$ PESs for the O + C$_2$ reaction. The solid curves correspond to trajectories starting from the bottom of the two entrance channels ($v=-0.5$), while the dashed ones  show the averaged excitation functions for the $(v=0,j=0)$ state. }
  \label{fig:xsecs_average}
\end{figure}

 Fig.~\ref{fig:xsecs_average} compares the excitation functions, $\sigma_{\mathrm{R}}(E_{\mathrm{col}})$, obtained for the two C$_2$ electronic states. To compute these total cross sections, the results from the three triplet PESs were averaged as follows:
\begin{equation}
\sigma_{\mathrm{R}}(v,j;E_{\mathrm{col}}) = \frac{1}{3}\sum_{\alpha} \sigma_{\mathrm{R}}^{\alpha}(v,j;E_{\mathrm{col}})
\end{equation}
where $\alpha$ denotes the $1\,^3A''$, $2\,^3A''$, and $1\,^3A'$ PESs. The $1/3$ factor accounts for the equal electronic statistical weights of the three surfaces correlating with the O$(^3\mathrm{P})$ + C$_2(\mathrm{X}\,^1\Sigma_{\mathrm{g}}^{+})$ entrance channel. Although the electronic degeneracy factor differs for the reaction involving C$_2(\mathrm{a}\,^3\Pi_{\mathrm{u}})$, the same scaling was applied to enable a direct comparison. This choice is justified because none of the remaining PESs correlating with the O$(^3\mathrm{P})$ + C$_2(\mathrm{a}\,^3\Pi_{\mathrm{u}})$ asymptote lead to electronic ground-state products. For each electronic state, the excitation functions are reported for the $(v=0, j=0)$ level and for an artificial $v=-0.5$ state, where the zero-point energy is removed and trajectories are initiated from the minimum of the corresponding C$_2$ potential well.

For $v=-0.5$ the reactivity is generally higher for the  C$_2(^3\Pi_{\mathrm{u}})$ state and only at the lowest collision energies is higher for C$_2(^1\Sigma_{\mathrm{g}}^{+})$, although the cross sections of both states tend to converge at high $E_{\mathrm{col}}$.  This higher cross section at very low $E_{\mathrm{col}}$ for trajectories initiated on the C$_2(^1\Sigma_{\mathrm{g}}^{+})$ state can be attributed to energetics: the asymptotic energy of C$_2(^1\Sigma_{\mathrm{g}}^{+})$ is slightly lower than that of  C$_2(^3\Pi_{\mathrm{u}})$ (see Fig. \ref{fig4}). On the other hand, the state-specific differences can be traced back to the geometries in the reactant valleys. The $^3\Pi_{\mathrm{u}}$ minimum corresponds to a slightly more extended C$_2$ bond length, which places the system closer to the attractive entrance valley of the global triplet PESs and facilitates access to the interaction well. Conversely, the $^1\Sigma_{\mathrm{g}}^{+}$ minimum lies at shorter internuclear distances, which impairs access to the global minimum since reaching it requires an elongation of the C–C bond. This geometrical effect is particularly pronounced on the $2\,^3A''$ PES because of the T-shaped barrier in the entrance channel, which also depends on $r_{\mathrm{CC}}$. 

When the zero-point vibrational energy is included, the differences observed for $v=-0.5$ disappear, and the two $v=0,j=0$ excitation functions are virtually identical. As discussed above for Fig.~\ref{fig:xsecs-panels}, the crossing between the two C$_2$ potential curves lies below the $v=0$ level, giving trajectories initiated from either electronic state access to the same classically allowed region of the adiabatic PESs. Interestingly, the $\sigma_{\mathrm{R}}(E_{\mathrm{col}})$ calculated for the $v=0, j=0$ state is typically closer to that obtained for the vibrationless C$_2(^1\Sigma_{\mathrm{g}}^{+})$ state. The exception occurs at the lowest collision energies, where the cross section instead resembles that of the vibrationless C$_2(^3\Pi_{\mathrm{u}})$ state.

Furthermore, comparing the artificial vibrationless limits with the $v=0, j=0$ states provides an estimate of the dynamical discrepancies that would be expected from a fully diabatic treatment. As discussed previously, computing a strict diabatic representation would require disentangling a dense manifold of electronic states, a task that remains computationally unaffordable. Since the dynamical differences between the true diabatic states would be significantly smaller than the discrepancies already observed between the extreme vibrationless cross sections, this comparison validates our adiabatic approximation.

\begin{figure}[ht!]
  \centering
  \includegraphics[width=1.0\linewidth]{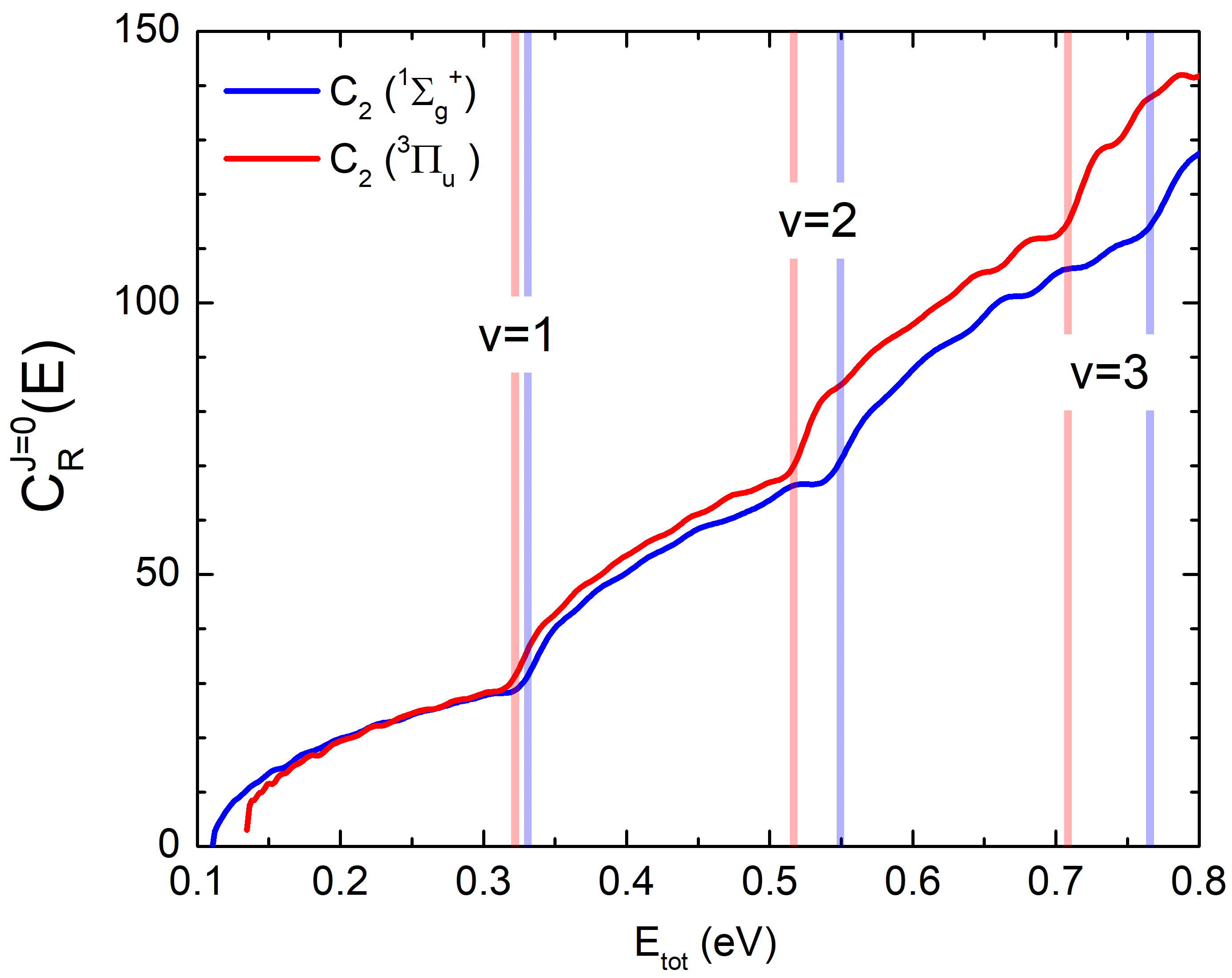}
  \caption{CRPs for $J$=0 as a function of the total energy  for the O + C$_2$($^1\Sigma_{\mathrm{g}}^{+}$) and the O + C$_2$($^3\Pi_\mathrm{u}$) reactions calculated on the $1\,^3A''$ PES. Energies are referred to the equilibrium geometry of the C$_2$($^1\Sigma_{\mathrm{g}}^{+}$) state. The opening of vibrational states for C$_2$($^1\Sigma_{\mathrm{g}}^{+}$) and C$_2$($^3\Pi_\mathrm{u}$) is shown as vertical blue and red lines, respectively.}
  \label{fig:crp}
\end{figure}

Finally, we calculated the cumulative reaction probabilities (CRPs) for total angular momentum $J=0$, which can be considered as a measure of the number
of reactant states that can proceed to products for a total energy. The CRP on the $1\,^3A''$ PES is shown in Fig.~\ref{fig:crp}. It increases monotonically with total energy, exhibiting a step-like structure that corresponds to the opening of successive C$_2$ vibrational channels, as observed for other barrierless reactions.\cite{Jambrina:JCP09, Jambrina:JCP12} Due to the high density of rotational states, individual rotational thresholds cannot be resolved. This step-like structure is indicative of reactions lacking strict dynamical bottlenecks; for reactions featuring narrow energetic valleys, the CRP reflects the continuous angular dependence of the barrier, which inherently blurs these sharp steps. Consequently, the step-like structure is much less pronounced on the $2\,^3A''$ PES (results not shown).

Beyond the macroscopic shape of the CRPs, which is consistent with the topography of the corresponding PESs, the most significant finding is the difference between the reactivity for the two C$_2$ states. While the CRP for the reaction with C$_2(^1\Sigma_{\mathrm{g}}^{+})$ is higher below the opening of the $v=1$ channel, it is surpassed by the C$_2(^3\Pi_{\mathrm{u}})$ state at higher vibrational energies. This crossover is a direct consequence of the higher density of states for the $^3\Pi_{\mathrm{u}}$ manifold, and the reactivity gap between the two states widens significantly as total energy increases. Accordingly, one would expect slightly larger thermal rate coefficients for the C$_2(^3\Pi_{\mathrm{u}})$ reaction at relatively high temperatures.

\section{Conclusions}\label{sec:conclusions}

In this work, we have computed three global adiabatic potential energy surfaces (PESs) for the O$(^3\mathrm{P})$ + C$_2(\mathrm{X}\,^1\Sigma_\mathrm{g}^+, \mathrm{a}\,^3\Pi_\mathrm{u})$ $\to$ C$(^3\mathrm{P})$ + CO$(\mathrm{X}\,^1\Sigma^+)$ reaction. The PESs were constructed from extensive icMRCI+Q calculations and fitted using a modified \texttt{GFIT3C} scheme, specifically adapted to describe the near-degenerate electronic structure of the entrance channel and the strong angular anisotropy present in the triatomic interaction region. The resulting analytical representations reproduce the \textit{ab initio} points with high accuracy over a broad configuration space, including both reactant and product asymptotes, the minima, and the saddle-point regions.

The three calculated PESs share several common characteristics. They all display an unusual double-well structure in the O + C$_2$ reactant valley, arising from the near degeneracy of the C$_2(^1\Sigma{_{\mathrm{g}}}^{+})$ and C$_2(^3\Pi_{\mathrm{u}})$ states and their adiabatic correlation with the same three triplet PESs. Moreover, all three surfaces are barrierless in the entrance channel and exhibit deep potential wells, partly caused by the strong exothermicity of the reaction. However, the PESs also exhibit marked differences. The $1\,^3A''$ surface is characterized by a single deep minimum corresponding to the collinear O--C--C geometry. In contrast, the $2\,^3A''$ and $1\,^3A'$ surfaces also exhibit collinear global minima, although they are considerably shallower, while remaining degenerate along the collinear $\Pi$ configuration and displaying submerged barriers in the exit channel.

Quasiclassical trajectory (QCT) calculations performed on the three analytical PESs confirm that the reaction proceeds without a barrier, in contrast to previous results. Furthermore, the excitation function decreases monotonically with increasing collision energy, as expected for a barrierless reaction featuring deep potential wells. The reaction cross sections are nearly identical regardless of whether the trajectories are initiated on the C$_2(^1\Sigma_{\mathrm{g}}^{+})$ or C$_2(^3\Pi_\mathrm{u})$ states. The only exception occurs when trajectories are run for an artificial $v=-0.5$ level (i.e., excluding zero-point vibrational energy); even in this case, the cross section is only slightly larger for the C$_2(^3\Pi_\mathrm{u})$ state. Cumulative reaction probabilities for $J=0$ reveal that the overall reactivity of the C$_2(^3\Pi_\mathrm{u})$ state is slightly higher at total energies above 0.3 eV due to its higher density of rovibrational states.

Overall, the present triplet PESs provide an accurate, internally consistent, and fully analytic description of the O + C$_2$ reactive system, overcoming the limitations of previous surfaces and offering improved reliability in both the entrance and product channels. Their accuracy and smoothness make them well-suited for future dynamical studies and for the modeling of astrophysical environments where O + C$_2$ collisions are relevant, particularly at the low collision energies characteristic of the interstellar medium.


\section*{ACKNOWLEDGMENTS}
 A. V. M. acknowledges grant no. EDU/1508/2020 (Junta de Castilla y León and European Social Fund). We acknowledge Grant No. PID 2023-147215NB-I00 and PID2024-155352NB-C22 funded by MCIN/AEI/10.13039/501100011033 Spanish Ministry of Science and Innovation, and FEDER, UE

\section*{AUTHOR DECLARATIONS}
\subsection*{Conflict of Interest}
The authors have no conflicts to disclose.

\subsection*{Author Contributions}
A. Veselinova:  Data Curation (lead); Investigation (lead); Formal analysis (equal); Conceptualization (equal). S. Gómez--Carrasco: Data Curation (equal); Methodology (equal); Software (equal); Conceptualization
(equal);. L González--Sánchez: Supervision (equal); Writing
– review \& editing (equal); Conceptualization
(equal). A. Zanchet: writing
– review \& editing (equal); Conceptualization
(equal); Software (equal). P. G. Jambrina: Conceptualization
(equal); Formal analysis (equal); Methodology (equal); Supervision
(equal); Writing – review \& editing (equal); Project administration (equal); Funding acquisition (lead). C. Sanz--Sanz: Data Curation (equal); Methodology (equal); Software (equal); Conceptualization (equal); Writing – review \& editing (equal).

\section*{DATA AVAILABILITY}
The ${1\,^3A''}$, ${2\,^3A''}$  and ${1\,^3A'}$  PESs computed in this article are openly available on Zenodo repository at https://doi.org/10.5281/zenodo.24320219.

\section*{REFERENCES}
\bibliography{cite}

\begin{thebibliography}{61}%
\makeatletter
\providecommand \@ifxundefined [1]{%
 \@ifx{#1\undefined}
}%
\providecommand \@ifnum [1]{%
 \ifnum #1\expandafter \@firstoftwo
 \else \expandafter \@secondoftwo
 \fi
}%
\providecommand \@ifx [1]{%
 \ifx #1\expandafter \@firstoftwo
 \else \expandafter \@secondoftwo
 \fi
}%
\providecommand \natexlab [1]{#1}%
\providecommand \enquote  [1]{``#1''}%
\providecommand \bibnamefont  [1]{#1}%
\providecommand \bibfnamefont [1]{#1}%
\providecommand \citenamefont [1]{#1}%
\providecommand \href@noop [0]{\@secondoftwo}%
\providecommand \href [0]{\begingroup \@sanitize@url \@href}%
\providecommand \@href[1]{\@@startlink{#1}\@@href}%
\providecommand \@@href[1]{\endgroup#1\@@endlink}%
\providecommand \@sanitize@url [0]{\catcode `\\12\catcode `\$12\catcode
  `\&12\catcode `\#12\catcode `\^12\catcode `\_12\catcode `\%12\relax}%
\providecommand \@@startlink[1]{}%
\providecommand \@@endlink[0]{}%
\providecommand \url  [0]{\begingroup\@sanitize@url \@url }%
\providecommand \@url [1]{\endgroup\@href {#1}{\urlprefix }}%
\providecommand \urlprefix  [0]{URL }%
\providecommand \Eprint [0]{\href }%
\providecommand \doibase [0]{https://doi.org/}%
\providecommand \selectlanguage [0]{\@gobble}%
\providecommand \bibinfo  [0]{\@secondoftwo}%
\providecommand \bibfield  [0]{\@secondoftwo}%
\providecommand \translation [1]{[#1]}%
\providecommand \BibitemOpen [0]{}%
\providecommand \bibitemStop [0]{}%
\providecommand \bibitemNoStop [0]{.\EOS\space}%
\providecommand \EOS [0]{\spacefactor3000\relax}%
\providecommand \BibitemShut  [1]{\csname bibitem#1\endcsname}%
\let\auto@bib@innerbib\@empty
\bibitem [{\citenamefont {{D. Troya}}\ and\ \citenamefont {{G. C.
  Schatz}}(2004)}]{troya_intrevphyschem_2004}%
  \BibitemOpen
  \bibfield  {author} {\bibinfo {author} {\bibnamefont {{D. Troya}}}\ and\
  \bibinfo {author} {\bibnamefont {{G. C. Schatz}}},\ }\bibfield  {title}
  {\enquote {\bibinfo {title} {{Hyperthermal chemistry in the gas phase and on
  surfaces: Theoretical studies}},}\ }\href
  {https://doi.org/10.1080/0144235042000298484} {\bibfield  {journal} {\bibinfo
   {journal} {Int. Rev. Phys. Chem.}\ }\textbf {\bibinfo {volume} {23(3)}},\
  \bibinfo {pages} {341} (\bibinfo {year} {2004})}\BibitemShut {NoStop}%
\bibitem [{\citenamefont {{A. Páramo}}\ \emph {et~al.}(2008)\citenamefont {{A.
  Páramo}}, \citenamefont {{A. Canosa}}, \citenamefont {{S. D. Le Picard}},\
  and\ \citenamefont {{I. R. Sims}}}]{paramo_jphyschema_2008}%
  \BibitemOpen
  \bibfield  {author} {\bibinfo {author} {\bibnamefont {{A. Páramo}}},
  \bibinfo {author} {\bibnamefont {{A. Canosa}}}, \bibinfo {author}
  {\bibnamefont {{S. D. Le Picard}}},\ and\ \bibinfo {author} {\bibnamefont
  {{I. R. Sims}}},\ }\bibfield  {title} {\enquote {\bibinfo {title} {{Rate
  coefficients for the reactions of C$_2(\mathrm{a}\, ^3\Pi_\mathrm{u})$ and
  C$_2(\mathrm{X}\,^1\Sigma^+_g)$ with various hydrocarbons (CH$_4$,
  C$_2$H$_2$, C$_2$H$_4$, C$_2$H$_6$, and C$_3$H$_8$): A gas-phase experimental
  study over the temperature range 24$-$300 K}},}\ }\href
  {https://doi.org/10.1021/jp8025336} {\bibfield  {journal} {\bibinfo
  {journal} {J. Phys. Chem. A}\ }\textbf {\bibinfo {volume} {112(39)}},\
  \bibinfo {pages} {9591} (\bibinfo {year} {2008})}\BibitemShut {NoStop}%
\bibitem [{\citenamefont {{X. Gu}}\ \emph {et~al.}(2006)\citenamefont {{X.
  Gu}}, \citenamefont {{Y. Guo}}, \citenamefont {{A. M. Mebel}},\ and\
  \citenamefont {{R. I. Kaiser}}}]{gu_jphyschema_2006}%
  \BibitemOpen
  \bibfield  {author} {\bibinfo {author} {\bibnamefont {{X. Gu}}}, \bibinfo
  {author} {\bibnamefont {{Y. Guo}}}, \bibinfo {author} {\bibnamefont {{A. M.
  Mebel}}},\ and\ \bibinfo {author} {\bibnamefont {{R. I. Kaiser}}},\
  }\bibfield  {title} {\enquote {\bibinfo {title} {{Chemical dynamics of the
  formation of the 1,3-butadiynyl radical (C$_4$H$(\mathrm{X}\,^2\Sigma^+)$)
  and its isotopomers}},}\ }\href {https://doi.org/10.1021/jp063256l}
  {\bibfield  {journal} {\bibinfo  {journal} {J. Phys. Chem. A}\ }\textbf
  {\bibinfo {volume} {110(39)}},\ \bibinfo {pages} {11265} (\bibinfo {year}
  {2006})}\BibitemShut {NoStop}%
\bibitem [{\citenamefont {{L. Pasternack}}, \citenamefont {{W. M. Pitt}},\ and\
  \citenamefont {{J. R. Mcdonald}}(1981)}]{pasternack_chemphys_1981}%
  \BibitemOpen
  \bibfield  {author} {\bibinfo {author} {\bibnamefont {{L. Pasternack}}},
  \bibinfo {author} {\bibnamefont {{W. M. Pitt}}},\ and\ \bibinfo {author}
  {\bibnamefont {{J. R. Mcdonald}}},\ }\bibfield  {title} {\enquote {\bibinfo
  {title} {{Temperature dependence of reactions and intersystem crossing of
  C$_2(\mathrm{a}\, ^3\Pi_\mathrm{u})$ with hydrogen and small hydrocarbons
  from 300-600 K}},}\ }\href {https://doi.org/10.1016/0301-0104(81)80017-1}
  {\bibfield  {journal} {\bibinfo  {journal} {Chem. Phys.}\ }\textbf {\bibinfo
  {volume} {57}},\ \bibinfo {pages} {19} (\bibinfo {year} {1981})}\BibitemShut
  {NoStop}%
\bibitem [{\citenamefont {{A. P. Baronavski}}\ and\ \citenamefont {{J. R.
  McDonald}}(1977)}]{barovavski_jchemphys_1977}%
  \BibitemOpen
  \bibfield  {author} {\bibinfo {author} {\bibnamefont {{A. P. Baronavski}}}\
  and\ \bibinfo {author} {\bibnamefont {{J. R. McDonald}}},\ }\bibfield
  {title} {\enquote {\bibinfo {title} {{Measurement of C$_2$ concentrations in
  an oxygen–acetylene flame: An application of saturation spectroscopy}},}\
  }\href {https://doi.org/10.1063/1.434306} {\bibfield  {journal} {\bibinfo
  {journal} {J. Chem. Phys.}\ }\textbf {\bibinfo {volume} {66(7)}},\ \bibinfo
  {pages} {3300} (\bibinfo {year} {1977})}\BibitemShut {NoStop}%
\bibitem [{\citenamefont {{H. Reisler}}, \citenamefont {{M. Mangir}},\ and\
  \citenamefont {{C. Wittig}}(1980)}]{reisler_chemphys_1980}%
  \BibitemOpen
  \bibfield  {author} {\bibinfo {author} {\bibnamefont {{H. Reisler}}},
  \bibinfo {author} {\bibnamefont {{M. Mangir}}},\ and\ \bibinfo {author}
  {\bibnamefont {{C. Wittig}}},\ }\bibfield  {title} {\enquote {\bibinfo
  {title} {{The kinetics of free radicals generated by IR laser photolysis. II.
  Reactions of C$_2(\mathrm{X}\, ^1\Sigma^+_g)$, C$_2(\mathrm{a}\,
  ^3\Pi_\mathrm{u})$, C$_3(\overline{\mathrm{X}}\,^1\Sigma^+_g)$ and CN$(X
  ^2\Sigma^+)$ with O$_2$}},}\ }\href
  {https://doi.org/10.1016/0301-0104(80)80019-X} {\bibfield  {journal}
  {\bibinfo  {journal} {Chem. Phys.}\ }\textbf {\bibinfo {volume} {47}},\
  \bibinfo {pages} {49} (\bibinfo {year} {1980})}\BibitemShut {NoStop}%
\bibitem [{\citenamefont {{A. Páramo}}\ \emph {et~al.}(2006)\citenamefont {{A.
  Páramo}}, \citenamefont {{A. Canosa}}, \citenamefont {{S. D. Le Picard}},\
  and\ \citenamefont {{I. R. Sims}}}]{paramo_jphyschema_2006}%
  \BibitemOpen
  \bibfield  {author} {\bibinfo {author} {\bibnamefont {{A. Páramo}}},
  \bibinfo {author} {\bibnamefont {{A. Canosa}}}, \bibinfo {author}
  {\bibnamefont {{S. D. Le Picard}}},\ and\ \bibinfo {author} {\bibnamefont
  {{I. R. Sims}}},\ }\bibfield  {title} {\enquote {\bibinfo {title} {{An
  experimental study of the intersystem crossing and reactions of
  C$_2(\mathrm{X}\, ^1\Sigma^+_g)$ and C$_2(\mathrm{a}\, ^3\Pi_\mathrm{u})$
  with O$_2$ and NO at very low temperature (24--300 K)}},}\ }\href
  {https://doi.org/10.1021/jp0544787} {\bibfield  {journal} {\bibinfo
  {journal} {J. Phys. Chem. A}\ }\textbf {\bibinfo {volume} {110(9)}},\
  \bibinfo {pages} {3121} (\bibinfo {year} {2006})}\BibitemShut {NoStop}%
\bibitem [{\citenamefont {{Y. Guo}}\ \emph {et~al.}(2006)\citenamefont {{Y.
  Guo}}, \citenamefont {{X. Gu}}, \citenamefont {{F. Zhang}}, \citenamefont
  {{A. M. Mebel}},\ and\ \citenamefont {{R. I.
  Kaiser}}}]{guo_aipconfproc_2006}%
  \BibitemOpen
  \bibfield  {author} {\bibinfo {author} {\bibnamefont {{Y. Guo}}}, \bibinfo
  {author} {\bibnamefont {{X. Gu}}}, \bibinfo {author} {\bibnamefont {{F.
  Zhang}}}, \bibinfo {author} {\bibnamefont {{A. M. Mebel}}},\ and\ \bibinfo
  {author} {\bibnamefont {{R. I. Kaiser}}},\ }\bibfield  {title} {\enquote
  {\bibinfo {title} {{Reaction dynamics of small carbon clusters with
  unsaturated hydrocarbons in the interstellar medium}},}\ }\href
  {https://doi.org/10.1063/1.2359538} {\bibfield  {journal} {\bibinfo
  {journal} {AIP Conf. Proc.}\ }\textbf {\bibinfo {volume} {855}},\ \bibinfo
  {pages} {42} (\bibinfo {year} {2006})}\BibitemShut {NoStop}%
\bibitem [{\citenamefont {{M. R. Combi}}\ and\ \citenamefont {{U.
  Fink}}(1997)}]{combi_astrophysj_1997}%
  \BibitemOpen
  \bibfield  {author} {\bibinfo {author} {\bibnamefont {{M. R. Combi}}}\ and\
  \bibinfo {author} {\bibnamefont {{U. Fink}}},\ }\bibfield  {title} {\enquote
  {\bibinfo {title} {{A critical study of molecular photodissociation and CHON
  grain sources for cometary C$_2$}},}\ }\href {https://doi.org/10.1086/304349}
  {\bibfield  {journal} {\bibinfo  {journal} {Astrophys. J.}\ }\textbf
  {\bibinfo {volume} {484(2)}},\ \bibinfo {pages} {879} (\bibinfo {year}
  {1997})}\BibitemShut {NoStop}%
\bibitem [{\citenamefont {{S. R. Federman}}\ and\ \citenamefont {{W. T.
  Huntress}}(1989)}]{federman_astrophysj_1989}%
  \BibitemOpen
  \bibfield  {author} {\bibinfo {author} {\bibnamefont {{S. R. Federman}}}\
  and\ \bibinfo {author} {\bibnamefont {{W. T. Huntress}}},\ }\bibfield
  {title} {\enquote {\bibinfo {title} {{Diffuse interstellar clouds as a
  chemical laboratory: The chemistry of diatomic carbon species}},}\ }\href
  {https://doi.org/10.1086/167187} {\bibfield  {journal} {\bibinfo  {journal}
  {Astrophys. J.}\ }\textbf {\bibinfo {volume} {338}},\ \bibinfo {pages} {140}
  (\bibinfo {year} {1989})}\BibitemShut {NoStop}%
\bibitem [{\citenamefont {{S. R. Federman}}\ and\ \citenamefont {{D. L.
  Lambert}}(1988)}]{federman_astrohpysj_1988}%
  \BibitemOpen
  \bibfield  {author} {\bibinfo {author} {\bibnamefont {{S. R. Federman}}}\
  and\ \bibinfo {author} {\bibnamefont {{D. L. Lambert}}},\ }\bibfield  {title}
  {\enquote {\bibinfo {title} {{The carbon chemistry in interstellar clouds
  toward moderately reddened stars}},}\ }\href {https://doi.org/10.1086/166336}
  {\bibfield  {journal} {\bibinfo  {journal} {Astrophys. J.}\ }\textbf
  {\bibinfo {volume} {328}},\ \bibinfo {pages} {777} (\bibinfo {year}
  {1988})}\BibitemShut {NoStop}%
\bibitem [{\citenamefont {{S. Green}}(1981)}]{green_annyrevphyschem_1981}%
  \BibitemOpen
  \bibfield  {author} {\bibinfo {author} {\bibnamefont {{S. Green}}},\
  }\bibfield  {title} {\enquote {\bibinfo {title} {{Interstellar chemistry:
  Exotic molecules in space}},}\ }\href
  {https://doi.org/10.1146/annurev.pc.32.100181.000535} {\bibfield  {journal}
  {\bibinfo  {journal} {Annu. Rev. Phys. Chem.}\ }\textbf {\bibinfo {volume}
  {32}},\ \bibinfo {pages} {103} (\bibinfo {year} {1981})}\BibitemShut
  {NoStop}%
\bibitem [{\citenamefont {{T. Kruse}}\ and\ \citenamefont {{P.
  Roth}}(1998)}]{kruse_sympintcomust_1998}%
  \BibitemOpen
  \bibfield  {author} {\bibinfo {author} {\bibnamefont {{T. Kruse}}}\ and\
  \bibinfo {author} {\bibnamefont {{P. Roth}}},\ }\bibfield  {title} {\enquote
  {\bibinfo {title} {{High-temperature reactions of C$_2$ with atomic and
  molecular oxygen}},}\ }\href {https://doi.org/10.1016/S0082-0784(98)80405-0}
  {\bibfield  {journal} {\bibinfo  {journal} {Symp. (Int.) Combust.}\ }\textbf
  {\bibinfo {volume} {27(1)}},\ \bibinfo {pages} {193} (\bibinfo {year}
  {1998})}\BibitemShut {NoStop}%
\bibitem [{\citenamefont {{H. Reisler}}, \citenamefont {{M. S. Mangir}},\ and\
  \citenamefont {{C. Wittig}}(1980)}]{reisler_jchemphys_1980}%
  \BibitemOpen
  \bibfield  {author} {\bibinfo {author} {\bibnamefont {{H. Reisler}}},
  \bibinfo {author} {\bibnamefont {{M. S. Mangir}}},\ and\ \bibinfo {author}
  {\bibnamefont {{C. Wittig}}},\ }\bibfield  {title} {\enquote {\bibinfo
  {title} {{Kinetics of free radicals generated by IR laser photolysis. IV.
  Intersystem crossings and reactions of C$_2(\mathrm{X}\, ^1\Sigma^+_g)$ and
  C$_2(\mathrm{a}\, ^3\Pi_\mathrm{u})$ in the gaseous phase}},}\ }\href
  {https://doi.org/10.1063/1.440377} {\bibfield  {journal} {\bibinfo  {journal}
  {J. Chem. Phys.}\ }\textbf {\bibinfo {volume} {73(5)}},\ \bibinfo {pages}
  {2280} (\bibinfo {year} {1980})}\BibitemShut {NoStop}%
\bibitem [{\citenamefont {{M. S. Mangir}}, \citenamefont {{H. Reisler}},\ and\
  \citenamefont {{C. Wittig}}(1980)}]{mangir_jchemphys_1980}%
  \BibitemOpen
  \bibfield  {author} {\bibinfo {author} {\bibnamefont {{M. S. Mangir}}},
  \bibinfo {author} {\bibnamefont {{H. Reisler}}},\ and\ \bibinfo {author}
  {\bibnamefont {{C. Wittig}}},\ }\bibfield  {title} {\enquote {\bibinfo
  {title} {{The kinetics of free radicals generated by IR laser photolysis.
  III. Intersystem crossing between C$_2(\mathrm{X}\, ^1\Sigma^+_g)$ and
  C$_2(\mathrm{a}\, ^3\Pi_\mathrm{u})$ induced by collisions with oxygen}},}\
  }\href {https://doi.org/10.1063/1.440190} {\bibfield  {journal} {\bibinfo
  {journal} {J. Chem. Phys.}\ }\textbf {\bibinfo {volume} {73(2)}},\ \bibinfo
  {pages} {829} (\bibinfo {year} {1980})}\BibitemShut {NoStop}%
\bibitem [{\citenamefont {{M. Kazmierczak}}\ \emph {et~al.}(2010)\citenamefont
  {{M. Kazmierczak}}, \citenamefont {{M. R. Schmidt}}, \citenamefont {{G. A.
  Galazutdinov}}, \citenamefont {{F. A. Musaev}}, \citenamefont {{Y.
  Betelesky}},\ and\ \citenamefont {{J. Krełowski}}}]{kazmierczak_mnras_2010}%
  \BibitemOpen
  \bibfield  {author} {\bibinfo {author} {\bibnamefont {{M. Kazmierczak}}},
  \bibinfo {author} {\bibnamefont {{M. R. Schmidt}}}, \bibinfo {author}
  {\bibnamefont {{G. A. Galazutdinov}}}, \bibinfo {author} {\bibnamefont {{F.
  A. Musaev}}}, \bibinfo {author} {\bibnamefont {{Y. Betelesky}}},\ and\
  \bibinfo {author} {\bibnamefont {{J. Krełowski}}},\ }\bibfield  {title}
  {\enquote {\bibinfo {title} {{Centrosymmetric molecules as possible carriers
  of diffuse interstellar bands}},}\ }\href
  {https://doi.org/10.1111/j.1365-2966.2010.17226.x} {\bibfield  {journal}
  {\bibinfo  {journal} {Mon. Not. R. Astron. Soc.}\ }\textbf {\bibinfo {volume}
  {408}},\ \bibinfo {pages} {1590} (\bibinfo {year} {2010})}\BibitemShut
  {NoStop}%
\bibitem [{\citenamefont {{Daniel E. Welty}}\ \emph {et~al.}(2013)\citenamefont
  {{Daniel E. Welty}}, \citenamefont {{J. Christopher Howk}}, \citenamefont
  {{Nicolas Lehner}},\ and\ \citenamefont {{John H.
  Black}}}]{welty_mnras_2013}%
  \BibitemOpen
  \bibfield  {author} {\bibinfo {author} {\bibnamefont {{Daniel E. Welty}}},
  \bibinfo {author} {\bibnamefont {{J. Christopher Howk}}}, \bibinfo {author}
  {\bibnamefont {{Nicolas Lehner}}},\ and\ \bibinfo {author} {\bibnamefont
  {{John H. Black}}},\ }\bibfield  {title} {\enquote {\bibinfo {title}
  {{Detection of interstellar C2 and C3 in the Small Magellanic Cloud}},}\
  }\href {https://doi.org/10.1093/mnras/sts093} {\bibfield  {journal} {\bibinfo
   {journal} {Mon. Not. R. Astron. Soc.}\ }\textbf {\bibinfo {volume} {428}},\
  \bibinfo {pages} {1107} (\bibinfo {year} {2013})}\BibitemShut {NoStop}%
\bibitem [{\citenamefont {{N. L. J. Cox}}\ and\ \citenamefont {{F.
  Patat}}(2014)}]{cox_aa_2014}%
  \BibitemOpen
  \bibfield  {author} {\bibinfo {author} {\bibnamefont {{N. L. J. Cox}}}\ and\
  \bibinfo {author} {\bibnamefont {{F. Patat}}},\ }\bibfield  {title} {\enquote
  {\bibinfo {title} {{Dense molecular clouds in the SN 2008fp host galaxy}},}\
  }\href {https://doi.org/10.1051/0004-6361/201219143} {\bibfield  {journal}
  {\bibinfo  {journal} {A\&A}\ }\textbf {\bibinfo {volume} {565}},\ \bibinfo
  {pages} {A61} (\bibinfo {year} {2014})}\BibitemShut {NoStop}%
\bibitem [{\citenamefont {Loison}\ \emph {et~al.}(2014)\citenamefont {Loison},
  \citenamefont {Wakelam}, \citenamefont {Hickson}, \citenamefont {Bergeat},\
  and\ \citenamefont {Mereau}}]{Loison:NNRAS114}%
  \BibitemOpen
  \bibfield  {author} {\bibinfo {author} {\bibfnamefont {J.-C.}\ \bibnamefont
  {Loison}}, \bibinfo {author} {\bibfnamefont {V.}~\bibnamefont {Wakelam}},
  \bibinfo {author} {\bibfnamefont {K.~M.}\ \bibnamefont {Hickson}}, \bibinfo
  {author} {\bibfnamefont {A.}~\bibnamefont {Bergeat}},\ and\ \bibinfo {author}
  {\bibfnamefont {R.}~\bibnamefont {Mereau}},\ }\bibfield  {title} {\enquote
  {\bibinfo {title} {The gas-phase chemistry of carbon chains in dark cloud
  chemical models},}\ }\href {https://doi.org/10.1093/mnras/stt1956} {\bibfield
   {journal} {\bibinfo  {journal} {MNRAS}\ }\textbf {\bibinfo {volume} {437}},\
  \bibinfo {pages} {930--945} (\bibinfo {year} {2014})}\BibitemShut {NoStop}%
\bibitem [{\citenamefont {{Ian W. M. Smith}}, \citenamefont {{Eric Herbst}},\
  and\ \citenamefont {{Qiang Chang}}(2004)}]{smith_mnras_2004}%
  \BibitemOpen
  \bibfield  {author} {\bibinfo {author} {\bibnamefont {{Ian W. M. Smith}}},
  \bibinfo {author} {\bibnamefont {{Eric Herbst}}},\ and\ \bibinfo {author}
  {\bibnamefont {{Qiang Chang}}},\ }\bibfield  {title} {\enquote {\bibinfo
  {title} {{Rapid neutral--neutral reactions at low temperatures: a new network
  and first results for TMC‐1}},}\ }\href
  {https://doi.org/10.1111/j.1365-2966.2004.07656.x} {\bibfield  {journal}
  {\bibinfo  {journal} {Mon. Not. R. Astron. Soc.}\ }\textbf {\bibinfo {volume}
  {350}},\ \bibinfo {pages} {323} (\bibinfo {year} {2004})}\BibitemShut
  {NoStop}%
\bibitem [{\citenamefont {{K. Kirby}}\ and\ \citenamefont {{B.
  Liu}}(1979)}]{kirby_jchemphys_1979}%
  \BibitemOpen
  \bibfield  {author} {\bibinfo {author} {\bibnamefont {{K. Kirby}}}\ and\
  \bibinfo {author} {\bibnamefont {{B. Liu}}},\ }\bibfield  {title} {\enquote
  {\bibinfo {title} {{The valence states of C$_2$: A configuration interaction
  study}},}\ }\href {https://doi.org/10.1063/1.437480} {\bibfield  {journal}
  {\bibinfo  {journal} {J. Chem. Phys.}\ }\textbf {\bibinfo {volume} {70}},\
  \bibinfo {pages} {893} (\bibinfo {year} {1979})}\BibitemShut {NoStop}%
\bibitem [{\citenamefont {{L. Veseth}}(1975)}]{veset_canjphys_1975}%
  \BibitemOpen
  \bibfield  {author} {\bibinfo {author} {\bibnamefont {{L. Veseth}}},\
  }\bibfield  {title} {\enquote {\bibinfo {title} {{Molecular Parameters for
  the $\mathrm{a}\,^3\Pi$ and $\mathrm{b}\,^3\Sigma_g^-$ States in C$_2$}},}\
  }\href {https://doi.org/10.1139/p75-039} {\bibfield  {journal} {\bibinfo
  {journal} {Can. J. Phys.}\ }\textbf {\bibinfo {volume} {53}},\ \bibinfo
  {pages} {299} (\bibinfo {year} {1975})}\BibitemShut {NoStop}%
\bibitem [{\citenamefont {{K. H. Becker}}\ \emph {et~al.}(2000)\citenamefont
  {{K. H. Becker}}, \citenamefont {{B. Donner}}, \citenamefont {{C. M. Freitas
  Dinis}}, \citenamefont {{H. Geiger}}, \citenamefont {{F. Schmidt}},\ and\
  \citenamefont {{P. Wiesen}}}]{becker_zphyschem_2000}%
  \BibitemOpen
  \bibfield  {author} {\bibinfo {author} {\bibnamefont {{K. H. Becker}}},
  \bibinfo {author} {\bibnamefont {{B. Donner}}}, \bibinfo {author}
  {\bibnamefont {{C. M. Freitas Dinis}}}, \bibinfo {author} {\bibnamefont {{H.
  Geiger}}}, \bibinfo {author} {\bibnamefont {{F. Schmidt}}},\ and\ \bibinfo
  {author} {\bibnamefont {{P. Wiesen}}},\ }\bibfield  {title} {\enquote
  {\bibinfo {title} {{Kinetics of the C$_2(\mathrm{a}\, ^3\Pi_\mathrm{u})$
  radical reacting with selected molecules and atoms}},}\ }\href
  {https://doi.org/10.1524/zpch.2000.214.4.503} {\bibfield  {journal} {\bibinfo
   {journal} {Z. Phys. Chem.}\ }\textbf {\bibinfo {volume} {214(4)}},\ \bibinfo
  {pages} {503} (\bibinfo {year} {2000})}\BibitemShut {NoStop}%
\bibitem [{\citenamefont {{M. Ray}}, \citenamefont {{B. Saha}},\ and\
  \citenamefont {{G. C. Schatz}}(2012)}]{schatz_jphyschemc_2012}%
  \BibitemOpen
  \bibfield  {author} {\bibinfo {author} {\bibnamefont {{M. Ray}}}, \bibinfo
  {author} {\bibnamefont {{B. Saha}}},\ and\ \bibinfo {author} {\bibnamefont
  {{G. C. Schatz}}},\ }\bibfield  {title} {\enquote {\bibinfo {title}
  {{Theoretical Studies of the O$(^3\mathrm{P})$ + C$_2$ Reaction at
  Hyperthermal Energies}},}\ }\href {https://doi.org/10.1021/jp3066629}
  {\bibfield  {journal} {\bibinfo  {journal} {J. Phys. Chem. C}\ }\textbf
  {\bibinfo {volume} {116(50)}},\ \bibinfo {pages} {26577} (\bibinfo {year}
  {2012})}\BibitemShut {NoStop}%
\bibitem [{\citenamefont {{D. Wang}}\ \emph {et~al.}(2019)\citenamefont {{D.
  Wang}}, \citenamefont {{D. Wang}}, \citenamefont {{L. Fu}},\ and\
  \citenamefont {{X. Huang}}}]{wang_chemphys_2019}%
  \BibitemOpen
  \bibfield  {author} {\bibinfo {author} {\bibnamefont {{D. Wang}}}, \bibinfo
  {author} {\bibnamefont {{D. Wang}}}, \bibinfo {author} {\bibnamefont {{L.
  Fu}}},\ and\ \bibinfo {author} {\bibnamefont {{X. Huang}}},\ }\bibfield
  {title} {\enquote {\bibinfo {title} {{Accurate potential surfaces for the
  first three lowest states of reaction O$(^3\mathrm{P})$ + C$_2(\mathrm{a}\,
  ^3\Pi_\mathrm{u})$ $\to$ CO$(\mathrm{X}\, ^1\Sigma)$ + C$(^1\mathrm{D})$}},}\
  }\href {https://doi.org/10.1016/j.chemphys.2018.09.013} {\bibfield  {journal}
  {\bibinfo  {journal} {Chem. Phys.}\ }\textbf {\bibinfo {volume} {517}},\
  \bibinfo {pages} {228} (\bibinfo {year} {2019})}\BibitemShut {NoStop}%
\bibitem [{\citenamefont {{X. Huang}}\ \emph {et~al.}(2025)\citenamefont {{X.
  Huang}}, \citenamefont {{G. Wang}}, \citenamefont {{C. Guo}}, \citenamefont
  {{X. Cheng}},\ and\ \citenamefont {{H. Zhang}}}]{zang_jchemphys_2015}%
  \BibitemOpen
  \bibfield  {author} {\bibinfo {author} {\bibnamefont {{X. Huang}}}, \bibinfo
  {author} {\bibnamefont {{G. Wang}}}, \bibinfo {author} {\bibnamefont {{C.
  Guo}}}, \bibinfo {author} {\bibnamefont {{X. Cheng}}},\ and\ \bibinfo
  {author} {\bibnamefont {{H. Zhang}}},\ }\bibfield  {title} {\enquote
  {\bibinfo {title} {{Reactive rate coefficients and machine learning
  predictions for O$(^3\mathrm{P})$ + C$_2(\mathrm{X}\, ^1\Sigma^+_g)$
  collisions on an accurate PIP-NN potential energy surface}},}\ }\href
  {https://doi.org/10.1063/5.0244191} {\bibfield  {journal} {\bibinfo
  {journal} {J. Chem. Phys.}\ }\textbf {\bibinfo {volume} {162}},\ \bibinfo
  {pages} {064307} (\bibinfo {year} {2025})}\BibitemShut {NoStop}%
\bibitem [{\citenamefont {Xu}\ \emph {et~al.}(2022)\citenamefont {Xu},
  \citenamefont {Federman}, \citenamefont {Jackson}, \citenamefont {Ng},
  \citenamefont {Wang},\ and\ \citenamefont {Crabtree}}]{Kyle:JCP22}%
  \BibitemOpen
  \bibfield  {author} {\bibinfo {author} {\bibfnamefont {Z.}~\bibnamefont
  {Xu}}, \bibinfo {author} {\bibfnamefont {S.~R.}\ \bibnamefont {Federman}},
  \bibinfo {author} {\bibfnamefont {W.~M.}\ \bibnamefont {Jackson}}, \bibinfo
  {author} {\bibfnamefont {C.-Y.}\ \bibnamefont {Ng}}, \bibinfo {author}
  {\bibfnamefont {L.-P.}\ \bibnamefont {Wang}},\ and\ \bibinfo {author}
  {\bibfnamefont {K.~N.}\ \bibnamefont {Crabtree}},\ }\bibfield  {title}
  {\enquote {\bibinfo {title} {{Multireference configuration interaction study
  of the predissociation of C$_2$ via its F$^1\Pi_u$ state}},}\ }\href
  {https://doi.org/10.1063/5.0097451} {\bibfield  {journal} {\bibinfo
  {journal} {J. Chem. Phys.}\ }\textbf {\bibinfo {volume} {157}},\ \bibinfo
  {pages} {024302} (\bibinfo {year} {2022})}\BibitemShut {NoStop}%
\bibitem [{\citenamefont {Rocha}\ \emph {et~al.}(2023)\citenamefont {Rocha},
  \citenamefont {Roncero}, \citenamefont {Bulut}, \citenamefont {Zuchowski},
  \citenamefont {Navarro-Almaida}, \citenamefont {Fuente}, \citenamefont
  {Wakelam}, \citenamefont {Loison}, \citenamefont {Roueff}, \citenamefont
  {Goicoechea}, \citenamefont {Esplugues}, \citenamefont {Beitia-Antero},
  \citenamefont {Caselli}, \citenamefont {Lattanzi}, \citenamefont {Pineda},
  \citenamefont {Le~Gal}, \citenamefont {Rodr\'{\i}guez-Baras},\ and\
  \citenamefont {Riviere-Marichalar}}]{OctavioSC2}%
  \BibitemOpen
  \bibfield  {author} {\bibinfo {author} {\bibfnamefont {C.~M.~R.}\
  \bibnamefont {Rocha}}, \bibinfo {author} {\bibfnamefont {O.}~\bibnamefont
  {Roncero}}, \bibinfo {author} {\bibfnamefont {N.}~\bibnamefont {Bulut}},
  \bibinfo {author} {\bibfnamefont {P.}~\bibnamefont {Zuchowski}}, \bibinfo
  {author} {\bibfnamefont {D.}~\bibnamefont {Navarro-Almaida}}, \bibinfo
  {author} {\bibfnamefont {A.}~\bibnamefont {Fuente}}, \bibinfo {author}
  {\bibfnamefont {V.}~\bibnamefont {Wakelam}}, \bibinfo {author} {\bibfnamefont
  {J.-C.}\ \bibnamefont {Loison}}, \bibinfo {author} {\bibfnamefont
  {E.}~\bibnamefont {Roueff}}, \bibinfo {author} {\bibfnamefont {J.~R.}\
  \bibnamefont {Goicoechea}}, \bibinfo {author} {\bibfnamefont
  {G.}~\bibnamefont {Esplugues}}, \bibinfo {author} {\bibfnamefont
  {L.}~\bibnamefont {Beitia-Antero}}, \bibinfo {author} {\bibfnamefont
  {P.}~\bibnamefont {Caselli}}, \bibinfo {author} {\bibfnamefont
  {V.}~\bibnamefont {Lattanzi}}, \bibinfo {author} {\bibfnamefont
  {J.}~\bibnamefont {Pineda}}, \bibinfo {author} {\bibfnamefont
  {R.}~\bibnamefont {Le~Gal}}, \bibinfo {author} {\bibfnamefont
  {M.}~\bibnamefont {Rodr\'{\i}guez-Baras}},\ and\ \bibinfo {author}
  {\bibfnamefont {P.}~\bibnamefont {Riviere-Marichalar}},\ }\bibfield  {title}
  {\enquote {\bibinfo {title} {Gas phase elemental abundances in molecular
  clouds ({GEMS}) - {VIII}. unlocking the {CS} chemistry: The {CH} + {S} $\to$
  {CS} + {H} and {C}$_2$ + {S} $\to$ {CS} + {C} reactions},}\ }\href
  {https://doi.org/10.1051/0004-6361/202346967} {\bibfield  {journal} {\bibinfo
   {journal} {A\&A}\ }\textbf {\bibinfo {volume} {677}},\ \bibinfo {pages}
  {A41} (\bibinfo {year} {2023})}\BibitemShut {NoStop}%
\bibitem [{\citenamefont {Werner}\ \emph {et~al.}(2020)\citenamefont {Werner},
  \citenamefont {Knowles}, \citenamefont {Manby}, \citenamefont {Black},
  \citenamefont {Doll}, \citenamefont {He{\ss}elmann}, \citenamefont {Kats},
  \citenamefont {Köhn}, \citenamefont {Korona}, \citenamefont {Kreplin},
  \citenamefont {Ma}, \citenamefont {Miller}, \citenamefont {Mitrushchenkov},
  \citenamefont {Peterson}, \citenamefont {Polyak}, \citenamefont {Rauhut},\
  and\ \citenamefont {Sibaev}}]{molpro2020}%
  \BibitemOpen
  \bibfield  {author} {\bibinfo {author} {\bibfnamefont {H.-J.}\ \bibnamefont
  {Werner}}, \bibinfo {author} {\bibfnamefont {P.~J.}\ \bibnamefont {Knowles}},
  \bibinfo {author} {\bibfnamefont {F.~R.}\ \bibnamefont {Manby}}, \bibinfo
  {author} {\bibfnamefont {J.~A.}\ \bibnamefont {Black}}, \bibinfo {author}
  {\bibfnamefont {K.}~\bibnamefont {Doll}}, \bibinfo {author} {\bibfnamefont
  {A.}~\bibnamefont {He{\ss}elmann}}, \bibinfo {author} {\bibfnamefont
  {D.}~\bibnamefont {Kats}}, \bibinfo {author} {\bibfnamefont {A.}~\bibnamefont
  {Köhn}}, \bibinfo {author} {\bibfnamefont {T.}~\bibnamefont {Korona}},
  \bibinfo {author} {\bibfnamefont {D.~A.}\ \bibnamefont {Kreplin}}, \bibinfo
  {author} {\bibfnamefont {Q.}~\bibnamefont {Ma}}, \bibinfo {author}
  {\bibfnamefont {T.~F.}\ \bibnamefont {Miller}}, \bibinfo {author}
  {\bibfnamefont {A.}~\bibnamefont {Mitrushchenkov}}, \bibinfo {author}
  {\bibfnamefont {K.~A.}\ \bibnamefont {Peterson}}, \bibinfo {author}
  {\bibfnamefont {I.}~\bibnamefont {Polyak}}, \bibinfo {author} {\bibfnamefont
  {G.}~\bibnamefont {Rauhut}},\ and\ \bibinfo {author} {\bibfnamefont
  {M.}~\bibnamefont {Sibaev}},\ }\bibfield  {title} {\enquote {\bibinfo {title}
  {The {Molpro} quantum chemistry package},}\ }\href
  {https://doi.org/10.1063/5.0005081} {\bibfield  {journal} {\bibinfo
  {journal} {J. Chem. Phys.}\ }\textbf {\bibinfo {volume} {152}},\ \bibinfo
  {pages} {144107} (\bibinfo {year} {2020})}\BibitemShut {NoStop}%
\bibitem [{\citenamefont {{T. H. Dunning, Jr.}}(1989)}]{dunning1}%
  \BibitemOpen
  \bibfield  {author} {\bibinfo {author} {\bibnamefont {{T. H. Dunning,
  Jr.}}},\ }\bibfield  {title} {\enquote {\bibinfo {title} {{Gaussian basis
  sets for use in correlated molecular calculations. I. The atoms boron through
  neon and hydrogen}},}\ }\href {https://doi.org/10.1063/1.456153} {\bibfield
  {journal} {\bibinfo  {journal} {J. Chem. Phys.}\ }\textbf {\bibinfo {volume}
  {90}},\ \bibinfo {pages} {1007} (\bibinfo {year} {1989})}\BibitemShut
  {NoStop}%
\bibitem [{\citenamefont {{R. A. Kendall}}, \citenamefont {{T. H. Dunning,
  Jr.}},\ and\ \citenamefont {{R. J. Harrison}}(1992)}]{dunning2}%
  \BibitemOpen
  \bibfield  {author} {\bibinfo {author} {\bibnamefont {{R. A. Kendall}}},
  \bibinfo {author} {\bibnamefont {{T. H. Dunning, Jr.}}},\ and\ \bibinfo
  {author} {\bibnamefont {{R. J. Harrison}}},\ }\bibfield  {title} {\enquote
  {\bibinfo {title} {{Electron affinities of the first‐row atoms revisited.
  Systematic basis sets and wave functions}},}\ }\href
  {https://doi.org/10.1063/1.462569} {\bibfield  {journal} {\bibinfo  {journal}
  {J. Chem. Phys.}\ }\textbf {\bibinfo {volume} {96}},\ \bibinfo {pages} {6796}
  (\bibinfo {year} {1992})}\BibitemShut {NoStop}%
\bibitem [{\citenamefont {{H.-J. Werner}}, \citenamefont {{B. Follmeg}},\ and\
  \citenamefont {Alexander}(1988)}]{mrci}%
  \BibitemOpen
  \bibfield  {author} {\bibinfo {author} {\bibnamefont {{H.-J. Werner}}},
  \bibinfo {author} {\bibnamefont {{B. Follmeg}}},\ and\ \bibinfo {author}
  {\bibfnamefont {M.~H.}\ \bibnamefont {Alexander}},\ }\bibfield  {title}
  {\enquote {\bibinfo {title} {{Adiabatic and diabatic potential energy
  surfaces for collisions of CN$(\mathrm{X}\,^2\Sigma^+, \mathrm{A}\,^2\Pi)$
  with He}},}\ }\href {https://doi.org/10.1063/1.454971} {\bibfield  {journal}
  {\bibinfo  {journal} {J. Chem. Phys.}\ }\textbf {\bibinfo {volume} {89}},\
  \bibinfo {pages} {3139} (\bibinfo {year} {1988})}\BibitemShut {NoStop}%
\bibitem [{\citenamefont {Davidson}(1975)}]{davidson}%
  \BibitemOpen
  \bibfield  {author} {\bibinfo {author} {\bibfnamefont {E.~R.}\ \bibnamefont
  {Davidson}},\ }\bibfield  {title} {\enquote {\bibinfo {title} {{The Iterative
  Calculation of a Few of the Lowest Eigenvalues and Corresponding Eigenvectors
  of Large Real-Symmetric Matrices}},}\ }\href
  {https://doi.org/10.1016/0021-9991(75)90065-0} {\bibfield  {journal}
  {\bibinfo  {journal} {J. Comput. Phys.}\ }\textbf {\bibinfo {volume} {17}},\
  \bibinfo {pages} {87} (\bibinfo {year} {1975})}\BibitemShut {NoStop}%
\bibitem [{\citenamefont {{G Verbockhaven}}\ \emph {et~al.}(2005)\citenamefont
  {{G Verbockhaven}}, \citenamefont {{C. Sanz Sanz}}, \citenamefont {{Gerrit C.
  Groenenboom}}, \citenamefont {{O. Roncero}},\ and\ \citenamefont {{Ad van der
  Avoird}}}]{gilles_jcp_2005}%
  \BibitemOpen
  \bibfield  {author} {\bibinfo {author} {\bibnamefont {{G Verbockhaven}}},
  \bibinfo {author} {\bibnamefont {{C. Sanz Sanz}}}, \bibinfo {author}
  {\bibnamefont {{Gerrit C. Groenenboom}}}, \bibinfo {author} {\bibnamefont
  {{O. Roncero}}},\ and\ \bibinfo {author} {\bibnamefont {{Ad van der
  Avoird}}},\ }\bibfield  {title} {\enquote {\bibinfo {title} {{Ab initio
  potential-energy surface for the reaction Ca + HCl $\to$ CaCl + H}},}\ }\href
  {https://doi.org/10.1063/1.1899154} {\bibfield  {journal} {\bibinfo
  {journal} {J. Chem. Phys.}\ }\textbf {\bibinfo {volume} {122}},\ \bibinfo
  {pages} {204307} (\bibinfo {year} {2005})}\BibitemShut {NoStop}%
\bibitem [{\citenamefont {{A. Aguado}}\ and\ \citenamefont {{M.
  Paniagua}}(1992)}]{fit1}%
  \BibitemOpen
  \bibfield  {author} {\bibinfo {author} {\bibnamefont {{A. Aguado}}}\ and\
  \bibinfo {author} {\bibnamefont {{M. Paniagua}}},\ }\bibfield  {title}
  {\enquote {\bibinfo {title} {{A new functional form to obtain analytical
  potentials of triatomic molecules}},}\ }\href
  {https://doi.org/10.1063/1.462163} {\bibfield  {journal} {\bibinfo  {journal}
  {J. Chem. Phys.}\ }\textbf {\bibinfo {volume} {96}},\ \bibinfo {pages} {1265}
  (\bibinfo {year} {1992})}\BibitemShut {NoStop}%
\bibitem [{\citenamefont {{A. Aguado}}, \citenamefont {{C. Suarez}},\ and\
  \citenamefont {{M. Paniagua}}(1993)}]{fit2}%
  \BibitemOpen
  \bibfield  {author} {\bibinfo {author} {\bibnamefont {{A. Aguado}}}, \bibinfo
  {author} {\bibnamefont {{C. Suarez}}},\ and\ \bibinfo {author} {\bibnamefont
  {{M. Paniagua}}},\ }\bibfield  {title} {\enquote {\bibinfo {title} {{Accurate
  fit of the two lowest excited‐state potential‐energy surfaces for doublet
  HeH$_2^+$}},}\ }\href {https://doi.org/10.1063/1.464676} {\bibfield
  {journal} {\bibinfo  {journal} {J. Chem. Phys.}\ }\textbf {\bibinfo {volume}
  {98}},\ \bibinfo {pages} {308} (\bibinfo {year} {1993})}\BibitemShut
  {NoStop}%
\bibitem [{\citenamefont {{A. Aguado}}, \citenamefont {{C. Tablero}},\ and\
  \citenamefont {{M. Paniagua}}(1998)}]{fit3}%
  \BibitemOpen
  \bibfield  {author} {\bibinfo {author} {\bibnamefont {{A. Aguado}}}, \bibinfo
  {author} {\bibnamefont {{C. Tablero}}},\ and\ \bibinfo {author} {\bibnamefont
  {{M. Paniagua}}},\ }\bibfield  {title} {\enquote {\bibinfo {title} {{Global
  fit of ab initio potential energy surfaces I. Triatomic systems}},}\ }\href
  {https://doi.org/10.1016/S0010-4655(97)00135-5} {\bibfield  {journal}
  {\bibinfo  {journal} {Comput. Phys. Commun.}\ }\textbf {\bibinfo {volume}
  {108}},\ \bibinfo {pages} {259} (\bibinfo {year} {1998})}\BibitemShut
  {NoStop}%
\bibitem [{\citenamefont {{Alexandre Zanchet}}\ \emph
  {et~al.}(2009)\citenamefont {{Alexandre Zanchet}}, \citenamefont {{Beatrice
  Bussery-Honvault}}, \citenamefont {{Mohamed Jorfib}},\ and\ \citenamefont
  {{Pascal Honvault}}}]{zanchet_pccp_2009}%
  \BibitemOpen
  \bibfield  {author} {\bibinfo {author} {\bibnamefont {{Alexandre Zanchet}}},
  \bibinfo {author} {\bibnamefont {{Beatrice Bussery-Honvault}}}, \bibinfo
  {author} {\bibnamefont {{Mohamed Jorfib}}},\ and\ \bibinfo {author}
  {\bibnamefont {{Pascal Honvault}}},\ }\bibfield  {title} {\enquote {\bibinfo
  {title} {{Study of the C$(^3\mathrm{P})$ + OH$(\mathrm{X}\,^2\Pi)$ $\to$
  CO$(\mathrm{a}\,^3\Pi)$ + H$(^2\mathrm{S})$ reaction: fully global ab initio
  potential energy surfaces of the $\mathrm{1}\,^2A''$ and $\mathrm{1}^4A''$
  excited states and non adiabatic couplings}},}\ }\href
  {https://doi.org/10.1039/B903829A} {\bibfield  {journal} {\bibinfo  {journal}
  {Phys. Chem. Chem. Phys.}\ }\textbf {\bibinfo {volume} {11}},\ \bibinfo
  {pages} {6182} (\bibinfo {year} {2009})}\BibitemShut {NoStop}%
\bibitem [{\citenamefont {{Susana Gómez-Carrasco}}, \citenamefont {{María Luz
  Hernández}},\ and\ \citenamefont {{Jose M.
  Alvariño}}(2007)}]{susana_chemphyslet_2007}%
  \BibitemOpen
  \bibfield  {author} {\bibinfo {author} {\bibnamefont {{Susana
  Gómez-Carrasco}}}, \bibinfo {author} {\bibnamefont {{María Luz
  Hernández}}},\ and\ \bibinfo {author} {\bibnamefont {{Jose M. Alvariño}}},\
  }\bibfield  {title} {\enquote {\bibinfo {title} {{Quantum and quasiclassical
  state-selected O($^1$D) + {HF} reaction dynamics and kinetics on a new MRCI
  ground singlet potential energy surface}},}\ }\href
  {https://doi.org/10.1016/j.cplett.2006.12.073} {\bibfield  {journal}
  {\bibinfo  {journal} {Chem. Phys Lett.}\ }\textbf {\bibinfo {volume} {435}},\
  \bibinfo {pages} {188} (\bibinfo {year} {2007})}\BibitemShut {NoStop}%
\bibitem [{\citenamefont {{J. Hernández-Rodríguez}}\ \emph
  {et~al.}(2023)\citenamefont {{J. Hernández-Rodríguez}}, \citenamefont {{C.
  Sanz-Sanz}}, \citenamefont {{P. Alberto Enríquez}}, \citenamefont {{M.
  González}},\ and\ \citenamefont {{M. Paniagua}}}]{jav_pccp_2023}%
  \BibitemOpen
  \bibfield  {author} {\bibinfo {author} {\bibnamefont {{J.
  Hernández-Rodríguez}}}, \bibinfo {author} {\bibnamefont {{C. Sanz-Sanz}}},
  \bibinfo {author} {\bibnamefont {{P. Alberto Enríquez}}}, \bibinfo {author}
  {\bibnamefont {{M. González}}},\ and\ \bibinfo {author} {\bibnamefont {{M.
  Paniagua}}},\ }\bibfield  {title} {\enquote {\bibinfo {title} {{Potential
  energy surfaces for singlet and triplet states of the LiH$_2^+$ system and
  quasi-classical trajectory cross sections for H + LiH$^+$ and H$^+$ +
  LiH}},}\ }\href {https://doi.org/10.1039/D3CP02959J} {\bibfield  {journal}
  {\bibinfo  {journal} {Phys. Chem. Chem. Phys.}\ }\textbf {\bibinfo {volume}
  {25}},\ \bibinfo {pages} {28052} (\bibinfo {year} {2023})}\BibitemShut
  {NoStop}%
\bibitem [{\citenamefont {{Alexandre Zanchet}}\ \emph
  {et~al.}(2019)\citenamefont {{Alexandre Zanchet}}, \citenamefont {{M.
  Menéndez}}, \citenamefont {{P. G. Jambrina}},\ and\ \citenamefont {{F.
  Javier Aoiz}}}]{zanchet_jcp_2019}%
  \BibitemOpen
  \bibfield  {author} {\bibinfo {author} {\bibnamefont {{Alexandre Zanchet}}},
  \bibinfo {author} {\bibnamefont {{M. Menéndez}}}, \bibinfo {author}
  {\bibnamefont {{P. G. Jambrina}}},\ and\ \bibinfo {author} {\bibnamefont {{F.
  Javier Aoiz}}},\ }\bibfield  {title} {\enquote {\bibinfo {title} {{New global
  potential energy surfaces of the ground $^3A'$ and $^3A''$ states of the
  O$(^3\mathrm{P})$ + H$_2$ system}},}\ }\href
  {https://doi.org/10.1063/1.5111844} {\bibfield  {journal} {\bibinfo
  {journal} {J. Chem. Phys.}\ }\textbf {\bibinfo {volume} {151}},\ \bibinfo
  {pages} {094307} (\bibinfo {year} {2019})}\BibitemShut {NoStop}%
\bibitem [{\citenamefont {{M. Baer}}(1975)}]{beaer_chphlett_1975}%
  \BibitemOpen
  \bibfield  {author} {\bibinfo {author} {\bibnamefont {{M. Baer}}},\
  }\bibfield  {title} {\enquote {\bibinfo {title} {{Adiabatic and diabatic
  representations for atom-molecule collisions: Treatment of the collinear
  arrangement}},}\ }\href {https://doi.org/10.1016/0009-2614(75)85599-0}
  {\bibfield  {journal} {\bibinfo  {journal} {Chem. Phys. Lett.}\ }\textbf
  {\bibinfo {volume} {35}},\ \bibinfo {pages} {112} (\bibinfo {year}
  {1975})}\BibitemShut {NoStop}%
\bibitem [{\citenamefont {Baer}(1980)}]{beaer_molphys_1980}%
  \BibitemOpen
  \bibfield  {author} {\bibinfo {author} {\bibfnamefont {M.}~\bibnamefont
  {Baer}},\ }\bibfield  {title} {\enquote {\bibinfo {title} {{Electronic
  non-adiabatic transitions derivation of the general adiabatic-diabatic
  transformation matrix}},}\ }\href {https://doi.org/10.1080/00268978000102091}
  {\bibfield  {journal} {\bibinfo  {journal} {Mol. Phys.}\ }\textbf {\bibinfo
  {volume} {40}},\ \bibinfo {pages} {1012} (\bibinfo {year}
  {1980})}\BibitemShut {NoStop}%
\bibitem [{\citenamefont {{A. Aguado}}\ \emph {et~al.}(2003)\citenamefont {{A.
  Aguado}}, \citenamefont {{M. Paniagua}}, \citenamefont {{C. Sanz Sanz}},\
  and\ \citenamefont {{O. Roncero}}}]{alfredo_jcp_2003_bis}%
  \BibitemOpen
  \bibfield  {author} {\bibinfo {author} {\bibnamefont {{A. Aguado}}}, \bibinfo
  {author} {\bibnamefont {{M. Paniagua}}}, \bibinfo {author} {\bibnamefont {{C.
  Sanz Sanz}}},\ and\ \bibinfo {author} {\bibnamefont {{O. Roncero}}},\
  }\bibfield  {title} {\enquote {\bibinfo {title} {{Transition state
  spectroscopy of the excited electronic states of Li--HF}},}\ }\href
  {https://doi.org/10.1063/1.1618223} {\bibfield  {journal} {\bibinfo
  {journal} {J. Chem. Phys.}\ }\textbf {\bibinfo {volume} {119}},\ \bibinfo
  {pages} {10088} (\bibinfo {year} {2003})}\BibitemShut {NoStop}%
\bibitem [{\citenamefont {{C. Sanz Sanz}}\ \emph {et~al.}(2002)\citenamefont
  {{C. Sanz Sanz}}, \citenamefont {{O. Roncero}}, \citenamefont {{M.
  Paniagua}},\ and\ \citenamefont {{A. Aguado}}}]{cris_chemphyslett_2002}%
  \BibitemOpen
  \bibfield  {author} {\bibinfo {author} {\bibnamefont {{C. Sanz Sanz}}},
  \bibinfo {author} {\bibnamefont {{O. Roncero}}}, \bibinfo {author}
  {\bibnamefont {{M. Paniagua}}},\ and\ \bibinfo {author} {\bibnamefont {{A.
  Aguado}}},\ }\bibfield  {title} {\enquote {\bibinfo {title} {{Transition
  state spectroscopy of the excited electronic states of Li--HF}},}\ }\href
  {https://doi.org/10.1016/S0009-2614(01)01387-2} {\bibfield  {journal}
  {\bibinfo  {journal} {Chem. Phys. Lett.}\ }\textbf {\bibinfo {volume}
  {351}},\ \bibinfo {pages} {295} (\bibinfo {year} {2002})}\BibitemShut
  {NoStop}%
\bibitem [{\citenamefont {{A. Aguado}}, \citenamefont {{O. Roncero}},\ and\
  \citenamefont {{C. Sanz Sanz}}(2021)}]{alfredo_pccp_2021}%
  \BibitemOpen
  \bibfield  {author} {\bibinfo {author} {\bibnamefont {{A. Aguado}}}, \bibinfo
  {author} {\bibnamefont {{O. Roncero}}},\ and\ \bibinfo {author} {\bibnamefont
  {{C. Sanz Sanz}}},\ }\bibfield  {title} {\enquote {\bibinfo {title} {{Three
  states global fittings with improved long range: singlet and triplet states
  of H$_3^+$}},}\ }\href {https://doi.org/10.1039/D0CP04100A} {\bibfield
  {journal} {\bibinfo  {journal} {Phys. Chem. Chem. Phys.}\ }\textbf {\bibinfo
  {volume} {23}},\ \bibinfo {pages} {7735} (\bibinfo {year}
  {2021})}\BibitemShut {NoStop}%
\bibitem [{\citenamefont {{A. Aguado}}\ \emph {et~al.}(2000)\citenamefont {{A.
  Aguado}}, \citenamefont {{O. Roncero}}, \citenamefont {{C. Tablero}},
  \citenamefont {{C. Sanz Sanz}},\ and\ \citenamefont {{M.
  Paniagua}}}]{alfredo_jcp_2000}%
  \BibitemOpen
  \bibfield  {author} {\bibinfo {author} {\bibnamefont {{A. Aguado}}}, \bibinfo
  {author} {\bibnamefont {{O. Roncero}}}, \bibinfo {author} {\bibnamefont {{C.
  Tablero}}}, \bibinfo {author} {\bibnamefont {{C. Sanz Sanz}}},\ and\ \bibinfo
  {author} {\bibnamefont {{M. Paniagua}}},\ }\bibfield  {title} {\enquote
  {\bibinfo {title} {{Global potential energy surfaces for the H$_3^+$ system.
  Analytical representation of the adiabatic ground-state $\mathrm{1}\,A'$
  potential}},}\ }\href {https://doi.org/10.1063/1.480539} {\bibfield
  {journal} {\bibinfo  {journal} {J. Chem. Phys.}\ }\textbf {\bibinfo {volume}
  {112}},\ \bibinfo {pages} {1240} (\bibinfo {year} {2000})}\BibitemShut
  {NoStop}%
\bibitem [{\citenamefont {{C. Sanz Sanz}}\ \emph {et~al.}(2001)\citenamefont
  {{C. Sanz Sanz}}, \citenamefont {{O. Roncero}}, \citenamefont {{C. Tablero}},
  \citenamefont {{A. Aguado}},\ and\ \citenamefont {{M.
  Paniagua}}}]{cris_jcp_2001}%
  \BibitemOpen
  \bibfield  {author} {\bibinfo {author} {\bibnamefont {{C. Sanz Sanz}}},
  \bibinfo {author} {\bibnamefont {{O. Roncero}}}, \bibinfo {author}
  {\bibnamefont {{C. Tablero}}}, \bibinfo {author} {\bibnamefont {{A.
  Aguado}}},\ and\ \bibinfo {author} {\bibnamefont {{M. Paniagua}}},\
  }\bibfield  {title} {\enquote {\bibinfo {title} {{The lowest triplet state
  $^3A'$ of H$_3^+$: Global potential energy surface and vibrational
  calculations}},}\ }\href {https://doi.org/10.1063/1.1336566} {\bibfield
  {journal} {\bibinfo  {journal} {J. Chem. Phys.}\ }\textbf {\bibinfo {volume}
  {114}},\ \bibinfo {pages} {2182} (\bibinfo {year} {2001})}\BibitemShut
  {NoStop}%
\bibitem [{\citenamefont {{C. Sanz Sanz}}\ \emph {et~al.}(2013)\citenamefont
  {{C. Sanz Sanz}}, \citenamefont {{O. Roncero}}, \citenamefont {{M.
  Paniagua}},\ and\ \citenamefont {{A. Aguado}}}]{cris_jcp_2013}%
  \BibitemOpen
  \bibfield  {author} {\bibinfo {author} {\bibnamefont {{C. Sanz Sanz}}},
  \bibinfo {author} {\bibnamefont {{O. Roncero}}}, \bibinfo {author}
  {\bibnamefont {{M. Paniagua}}},\ and\ \bibinfo {author} {\bibnamefont {{A.
  Aguado}}},\ }\bibfield  {title} {\enquote {\bibinfo {title} {{Full
  dimensional potential energy surface for the ground state of H$_4^+$ system
  based on triatomic-in-molecules formalism}},}\ }\href
  {https://doi.org/10.1063/1.4827640} {\bibfield  {journal} {\bibinfo
  {journal} {J. Chem. Phys.}\ }\textbf {\bibinfo {volume} {139}},\ \bibinfo
  {pages} {184302} (\bibinfo {year} {2013})}\BibitemShut {NoStop}%
\bibitem [{\citenamefont {{F. J. Aoiz}}, \citenamefont {{V. J. Herrero}},\ and\
  \citenamefont {{V. S\'aez-R\'abanos}}(1992)}]{qct1}%
  \BibitemOpen
  \bibfield  {author} {\bibinfo {author} {\bibnamefont {{F. J. Aoiz}}},
  \bibinfo {author} {\bibnamefont {{V. J. Herrero}}},\ and\ \bibinfo {author}
  {\bibnamefont {{V. S\'aez-R\'abanos}}},\ }\bibfield  {title} {\enquote
  {\bibinfo {title} {{Quasiclassical state to state reaction cross sections for
  D+H$_2$(v=0, j=0)$\rightarrow$HD(v',j')+H. Formation and characteristics of
  short-lived collision complexes}},}\ }\href
  {https://doi.org/10.1063/1.463514} {\bibfield  {journal} {\bibinfo  {journal}
  {J. Chem. Phys.}\ }\textbf {\bibinfo {volume} {97}},\ \bibinfo {pages} {7423}
  (\bibinfo {year} {1992})}\BibitemShut {NoStop}%
\bibitem [{\citenamefont {{F. J. Aoiz}}, \citenamefont {{L. Ba\~{n}ares}},\
  and\ \citenamefont {{V. J. Herrero}}(1998)}]{qct2}%
  \BibitemOpen
  \bibfield  {author} {\bibinfo {author} {\bibnamefont {{F. J. Aoiz}}},
  \bibinfo {author} {\bibnamefont {{L. Ba\~{n}ares}}},\ and\ \bibinfo {author}
  {\bibnamefont {{V. J. Herrero}}},\ }\bibfield  {title} {\enquote {\bibinfo
  {title} {{Recent results from quasiclassical trajectory computations of
  elementary chemical reactions}},}\ }\href {https://doi.org/10.1039/A803469I}
  {\bibfield  {journal} {\bibinfo  {journal} {J. Chem. Soc. Faraday. Trans.}\
  }\textbf {\bibinfo {volume} {94}},\ \bibinfo {pages} {2483} (\bibinfo {year}
  {1998})}\BibitemShut {NoStop}%
\bibitem [{\citenamefont {{F. J. Aoiz}}\ \emph {et~al.}(2006)\citenamefont {{F.
  J. Aoiz}}, \citenamefont {{M. Brouard}}, \citenamefont {{C. J. Eyles}},
  \citenamefont {{J. F. Castillo}},\ and\ \citenamefont {{V.
  S\'aez-R\'abanos}}}]{Aoiz:JCP06}%
  \BibitemOpen
  \bibfield  {author} {\bibinfo {author} {\bibnamefont {{F. J. Aoiz}}},
  \bibinfo {author} {\bibnamefont {{M. Brouard}}}, \bibinfo {author}
  {\bibnamefont {{C. J. Eyles}}}, \bibinfo {author} {\bibnamefont {{J. F.
  Castillo}}},\ and\ \bibinfo {author} {\bibnamefont {{V. S\'aez-R\'abanos}}},\
  }\bibfield  {title} {\enquote {\bibinfo {title} {{Cumulative reaction
  probabilities: A comparison between quasiclassical and quantum mechanical
  results}},}\ }\href {https://doi.org/10.1063/1.2353837} {\bibfield  {journal}
  {\bibinfo  {journal} {J. Chem. Phys.}\ }\textbf {\bibinfo {volume} {125}},\
  \bibinfo {pages} {144105} (\bibinfo {year} {2006})}\BibitemShut {NoStop}%
\bibitem [{\citenamefont {{P. G. Jambrina}}\ \emph {et~al.}(2009)\citenamefont
  {{P. G. Jambrina}}, \citenamefont {{F. J. Aoiz}}, \citenamefont {{C. J.
  Eyles}}, \citenamefont {{V. J. Herrero}},\ and\ \citenamefont {{V.
  S\'aez-R\'abanos}}}]{Jambrina:JCP09}%
  \BibitemOpen
  \bibfield  {author} {\bibinfo {author} {\bibnamefont {{P. G. Jambrina}}},
  \bibinfo {author} {\bibnamefont {{F. J. Aoiz}}}, \bibinfo {author}
  {\bibnamefont {{C. J. Eyles}}}, \bibinfo {author} {\bibnamefont {{V. J.
  Herrero}}},\ and\ \bibinfo {author} {\bibnamefont {{V. S\'aez-R\'abanos}}},\
  }\bibfield  {title} {\enquote {\bibinfo {title} {{Cumulative reaction
  probabilities and transition state properties: A study of the H$^+$+H$_2$ and
  H$^+$+D$_2$ proton exchange reactions}},}\ }\href
  {https://doi.org/10.1063/1.3129343} {\bibfield  {journal} {\bibinfo
  {journal} {J. Chem. Phys.}\ }\textbf {\bibinfo {volume} {130}},\ \bibinfo
  {pages} {184303} (\bibinfo {year} {2009})}\BibitemShut {NoStop}%
\bibitem [{\citenamefont {Truhlar}\ and\ \citenamefont
  {Muckerman}(1979)}]{truhlar79}%
  \BibitemOpen
  \bibfield  {author} {\bibinfo {author} {\bibfnamefont {D.~G.}\ \bibnamefont
  {Truhlar}}\ and\ \bibinfo {author} {\bibfnamefont {J.~T.}\ \bibnamefont
  {Muckerman}},\ }\href@noop {} {\emph {\bibinfo {title} {Reactive Scattering
  Cross Sections: Quasiclassical and Semiclassical Methods in Atom-Molecule
  Collision Theory: A Guide of the Experimentalists (Edited by R. B.
  Bernstein)}}}\ (\bibinfo  {publisher} {Plenum, New York},\ \bibinfo {year}
  {1979})\ pp.\ \bibinfo {pages} {505--566}\BibitemShut {NoStop}%
\bibitem [{\citenamefont {{Jeffery S. Boschen}}\ \emph
  {et~al.}(2014)\citenamefont {{Jeffery S. Boschen}}, \citenamefont {{Daniel
  Theis}}, \citenamefont {{Klaus Ruedenberg}},\ and\ \citenamefont {{Theresa L.
  Windus}}}]{boschen_theochemacc_2014}%
  \BibitemOpen
  \bibfield  {author} {\bibinfo {author} {\bibnamefont {{Jeffery S. Boschen}}},
  \bibinfo {author} {\bibnamefont {{Daniel Theis}}}, \bibinfo {author}
  {\bibnamefont {{Klaus Ruedenberg}}},\ and\ \bibinfo {author} {\bibnamefont
  {{Theresa L. Windus}}},\ }\bibfield  {title} {\enquote {\bibinfo {title}
  {{Accurate ab initio potential energy curves and spectroscopic properties of
  the four lowest singlet states of C$_2$}},}\ }\href
  {https://doi.org/10.1007/s00214-013-1425-x} {\bibfield  {journal} {\bibinfo
  {journal} {Theor. Chem. Acc.}\ }\textbf {\bibinfo {volume} {133}},\ \bibinfo
  {pages} {1425} (\bibinfo {year} {2014})}\BibitemShut {NoStop}%
\bibitem [{\citenamefont {{Lu Peng-Fei}}\ \emph {et~al.}(2013)\citenamefont
  {{Lu Peng-Fei}}, \citenamefont {{Yan Lei}}, \citenamefont {{Yu Zhong-Yuan}},
  \citenamefont {{Gao Yu-Feng}},\ and\ \citenamefont {{Gao
  Tao}}}]{lu_commtheophys_2013}%
  \BibitemOpen
  \bibfield  {author} {\bibinfo {author} {\bibnamefont {{Lu Peng-Fei}}},
  \bibinfo {author} {\bibnamefont {{Yan Lei}}}, \bibinfo {author} {\bibnamefont
  {{Yu Zhong-Yuan}}}, \bibinfo {author} {\bibnamefont {{Gao Yu-Feng}}},\ and\
  \bibinfo {author} {\bibnamefont {{Gao Tao}}},\ }\bibfield  {title} {\enquote
  {\bibinfo {title} {{An Accurate Calculation of Potential Energy Curves and
  Transition Dipole Moment for Low-Lying Electronic States of CO}},}\ }\href
  {https://doi.org/10.1088/0253-6102/59/2/11} {\bibfield  {journal} {\bibinfo
  {journal} {Commun. Theor. Phys.}\ }\textbf {\bibinfo {volume} {59}},\
  \bibinfo {pages} {193} (\bibinfo {year} {2013})}\BibitemShut {NoStop}%
\bibitem [{\citenamefont {{R. Pino-Rios}}\ and\ \citenamefont {{O.
  Denis-Alpizar}}(2023)}]{pinorios_intjquantumchem_2023}%
  \BibitemOpen
  \bibfield  {author} {\bibinfo {author} {\bibnamefont {{R. Pino-Rios}}}\ and\
  \bibinfo {author} {\bibnamefont {{O. Denis-Alpizar}}},\ }\bibfield  {title}
  {\enquote {\bibinfo {title} {{Two three-electron bonds in two triatomic
  molecules? The cases of C$_2$O and N$_2$C in their ground states}},}\ }\href
  {https://doi.org/10.1002/qua.27257} {\bibfield  {journal} {\bibinfo
  {journal} {Int. J. Quantum Chem.}\ }\textbf {\bibinfo {volume} {124(1)}},\
  \bibinfo {pages} {e27257} (\bibinfo {year} {2023})}\BibitemShut {NoStop}%
\bibitem [{\citenamefont {{V. Zengin}}\ \emph {et~al.}(1996)\citenamefont {{V.
  Zengin}}, \citenamefont {{B. Joakim Persson}}, \citenamefont {{K. M.
  Strong}},\ and\ \citenamefont {{R. E. Continetti}}}]{zengin_jchempyhs_1996}%
  \BibitemOpen
  \bibfield  {author} {\bibinfo {author} {\bibnamefont {{V. Zengin}}}, \bibinfo
  {author} {\bibnamefont {{B. Joakim Persson}}}, \bibinfo {author}
  {\bibnamefont {{K. M. Strong}}},\ and\ \bibinfo {author} {\bibnamefont {{R.
  E. Continetti}}},\ }\bibfield  {title} {\enquote {\bibinfo {title} {{Study of
  the low-lying electronic states of CCO by photoelectron spectroscopy of
  CCO$^-$ and ab initio calculations}},}\ }\href
  {https://doi.org/10.1063/1.473000} {\bibfield  {journal} {\bibinfo  {journal}
  {J. Chem. Phys.}\ }\textbf {\bibinfo {volume} {105(22)}},\ \bibinfo {pages}
  {9740} (\bibinfo {year} {1996})}\BibitemShut {NoStop}%
\bibitem [{\citenamefont {{Y. Ohshima}}, \citenamefont {{Y. Endo}},\ and\
  \citenamefont {{T. Ogata}}(1995)}]{ohshima_jchemphys_1995}%
  \BibitemOpen
  \bibfield  {author} {\bibinfo {author} {\bibnamefont {{Y. Ohshima}}},
  \bibinfo {author} {\bibnamefont {{Y. Endo}}},\ and\ \bibinfo {author}
  {\bibnamefont {{T. Ogata}}},\ }\bibfield  {title} {\enquote {\bibinfo {title}
  {{Fourier-transform microwave spectroscopy of triplet carbon monoxides,
  C$_2$O, C$_4$O, C$_6$O, and C$_8$O}},}\ }\href
  {https://doi.org/10.1063/1.468881} {\bibfield  {journal} {\bibinfo  {journal}
  {J. Chem. Phys.}\ }\textbf {\bibinfo {volume} {102(4)}},\ \bibinfo {pages}
  {1493} (\bibinfo {year} {1995})}\BibitemShut {NoStop}%
\bibitem [{\citenamefont {{H. Choi}}\ \emph {et~al.}(1998)\citenamefont {{H.
  Choi}}, \citenamefont {{D. H. Mordaunt}}, \citenamefont {{R. T. Bise}},
  \citenamefont {{T. R. Taylor}},\ and\ \citenamefont {{D. M.
  Neumark}}}]{choi_jchemphys_1998}%
  \BibitemOpen
  \bibfield  {author} {\bibinfo {author} {\bibnamefont {{H. Choi}}}, \bibinfo
  {author} {\bibnamefont {{D. H. Mordaunt}}}, \bibinfo {author} {\bibnamefont
  {{R. T. Bise}}}, \bibinfo {author} {\bibnamefont {{T. R. Taylor}}},\ and\
  \bibinfo {author} {\bibnamefont {{D. M. Neumark}}},\ }\bibfield  {title}
  {\enquote {\bibinfo {title} {{Photodissociation of triplet and singlet states
  of the CCO radical}},}\ }\href {https://doi.org/10.1063/1.475839} {\bibfield
  {journal} {\bibinfo  {journal} {J. Chem. Phys.}\ }\textbf {\bibinfo {volume}
  {108(10)}},\ \bibinfo {pages} {4070} (\bibinfo {year} {1998})}\BibitemShut
  {NoStop}%
\bibitem [{\citenamefont {Jambrina}\ \emph {et~al.}(2012)\citenamefont
  {Jambrina}, \citenamefont {Lara}, \citenamefont {Men\'endez}, \citenamefont
  {Launay},\ and\ \citenamefont {Aoiz}}]{Jambrina:JCP12}%
  \BibitemOpen
  \bibfield  {author} {\bibinfo {author} {\bibfnamefont {P.~G.}\ \bibnamefont
  {Jambrina}}, \bibinfo {author} {\bibfnamefont {M.}~\bibnamefont {Lara}},
  \bibinfo {author} {\bibfnamefont {M.}~\bibnamefont {Men\'endez}}, \bibinfo
  {author} {\bibfnamefont {J.-M.}\ \bibnamefont {Launay}},\ and\ \bibinfo
  {author} {\bibfnamefont {F.~J.}\ \bibnamefont {Aoiz}},\ }\bibfield  {title}
  {\enquote {\bibinfo {title} {Rate coefficients from quantum and
  quasi-classical cumulative reaction probabilities for the {S}($^1${D}) +
  {H}$_2$ reaction},}\ }\href {https://doi.org/10.1063/1.4761894} {\bibfield
  {journal} {\bibinfo  {journal} {J. Chem. Phys.}\ }\textbf {\bibinfo {volume}
  {137}},\ \bibinfo {pages} {164314} (\bibinfo {year} {2012})}\BibitemShut
  {NoStop}%
\end{thebibliography}%

\end{document}